\documentclass[twocolumn]{aastex63}
\usepackage{chemformula}
\usepackage{amsmath,amssymb,CJK}
\let\ce\ch
\usepackage[caption=false]{subfig}
\makeatletter
\g@addto@macro{\UrlBreaks}{\UrlOrds}
\makeatother
\PassOptionsToPackage{hyphens}{url}\usepackage{hyperref}
\urldef{\footurl}\url{https://news.arizona.edu/story/ua-researcher-captures-rare-radar-images-comet-46pwirtanen}
\usepackage{soul, CJK}

\definecolor{medium-blue}{rgb}{0,0,1}
\hypersetup{colorlinks, urlcolor={medium-blue}}
\usepackage{graphicx}

\received{September 29, 2020}
\revised{November 23, 2020}
\accepted{December 1, 2020}
\submitjournal{Planetary Science Journal}

\shorttitle{HST COS Observations of 46P}
\shortauthors{Noonan et al.}

\begin{document}
\begin{CJK*}{UTF8}{gbsn}
\title{FUV Observations of the Inner Coma of 46P/Wirtanen}
\correspondingauthor{John W. Noonan}
\email{noonan@lpl.arizona.edu}

\author[0000-0003-2152-6987]{John W. Noonan}
\affiliation{Lunar and Planetary Laboratory,University of Arizona, 1629 E University Blvd, Tucson, Arizona 85721-0092, USA }

\author{Walter M. Harris}
\affiliation{Lunar and Planetary Laboratory,University of Arizona, 1629 E University Blvd, Tucson, Arizona 85721-0092, USA }

\author[0000-0003-2110-8152]{Steven Bromley}
\affiliation{Physics Department, Leach Science Center, Auburn University, Auburn, AL 36849, USA}

\author[0000-0003-0774-884X]{Davide Farnocchia}
\affiliation{Jet Propulsion Laboratory, California Institute of Technology, 4800 Oak Grove Drive, Pasadena, 91109 CA, USA}

\author[0000-0003-3841-9977]{Jian-Yang Li (李荐扬)}
\affiliation{Planetary Science Institute, 1700 E. Ft. Lowell Rd., Suite 106, Tucson, AZ 85719, USA}

\author{Kathleen E. Mandt}
\affiliation{Johns Hopkins Applied Physics Laboratory, 11100 Johns Hopkins Rd., Laurel, MD, 20723, USA}

\author[0000-0002-3672-0603]{Joel Wm. Parker}
\affiliation{Planetary Science Directorate, Southwest Research Institute, Suite 300, 1050 Walnut Street, Boulder, Colorado 80302,USA}

\author[0000-0003-3321-1472]{Kumar Venkataramani}
\affiliation{Physics Department, Leach Science Center, Auburn University, Auburn, AL 36849, USA}

\author[0000-0002-2668-7248]{Dennis Bodewits}
\affiliation{Physics Department, Leach Science Center, Auburn University, Auburn, AL 36849, USA}

\begin{abstract}
Far ultraviolet observations of comets yield information about the energetic processes that dissociate the sublimated gases from their primitive surfaces. Understanding which emission processes are dominant, their effects on the observed cometary spectrum, and how to properly invert the spectrum back to composition of the presumably pristine surface ices of a comet nuclei are all critical components for proper interpretation and analysis of comets. The close approach of comet 46P/Wirtanen in 2018-2019 provided a unique opportunity to study the inner most parts of a cometary coma with the Hubble Space Telescope Cosmic Origins Spectrograph, rarely accessible with remote observations, at length scales (100's of km) and wavelengths (900-1430 \AA\~) previously probed only by the European Space Agency's \emph{Rosetta} spacecraft. Our observations show a complex picture for the inner coma; atomic production rates for H and O that show water is the dominant source of both, an abundance of atomic sulfur that is difficult to explain with the lifetimes of common sulfur parent molecules, and a density distribution that is poorly fit with both Haser and vectorial models.
\end{abstract}

\section{Introduction}\label{Intro}

Comets are frequently described as the well-preserved primitive remnants of our solar system's formation, offering insight into the chemical abundances of their formation location in the protoplanetary disk. This assumption is based on the source regions for comets: the Kuiper-Edgeworth Belt for Jupiter-family comets (JFCs) and the Oort Cloud for long-period comets (LPCs), both of which have temperatures well below the sublimation point of common volatiles and should preserve the primordial ice abundances over billion year timescales \citep{ahearn_comets_2011}. Directly measuring the ice compositions of comet nuclei is nearly impossible, unless icy grains are lifted from the surface and can be studied (e.g. \citet{protopapa_icy_2018}), and we are limited to investigating the emissions from gas in the coma produced by sublimating cometary ices. These emissions require modeling of the distribution of molecules and emission mechanisms in the coma to properly derive abundance of the parent molecules and tie them to nucleus properties \citep{2004come.book..425F}. Far Ultraviolet (FUV) observations of cometary comae have been instrumental in characterizing the atomic budget of comets \citep{feldman_fuv_2017,bodewits_carbon_2020}, and provide a directly comparable metric to protoplanetary disks. However, our understanding of the physical processes that drive cometary emissions, and that are subsequently used to derive these primordial properties, is not yet complete. The European Space Agency's \textit{Rosetta} mission revealed the prevalence of dissociative electron impact in the coma of 67P/Churyumov-Gerasimenko at heliocentric distances beyond 2.8 au  in the near nucleus coma ($\leq$ 100 km), which can produce different spectral signatures than the more commonly detected resonance fluorescence and photodissociation mechanisms, and thus affects interpretation of those observations \citep{feldman_measurements_2015, bodewits2016changes}. The presence of dissociative electron impact could allow detection of molecules which do not have emission features, like \ce{O2}, in the extreme inner coma, as it produces direct emission from dissociation of parent molecules with unique emission line ratios. However, the process was most consistent at large heliocentric distances, and was more related to transient events nearer perihelion.  Such a process helped reveal the presence of faint water emissions from Europa attributed to a plume \citep{roth2017detection} and the presence of species that are difficult to detect via traditional photodissociation and fluorescence, as with \ce{O2} on both Ganymede and Callisto \citep{hall1998far,cunningham_detection_2015}.  This discovery brought forth two particular questions: Is dissociative electron impact present in other comet comae? What local conditions need to be met to produce significant dissociative electron impact emissions?

The unique combination of extensive prior observation, substantial contemporaneous observation during the 2018-2019 close approach, and the ability to plan observations implementing the lessons learned from the \textit{Rosetta} mission provided an unprecedented opportunity to characterize the atomic and molecular emissions of 46P in the FUV. 

\begin{figure*}
    \centering
    \includegraphics[width=0.7\textwidth]{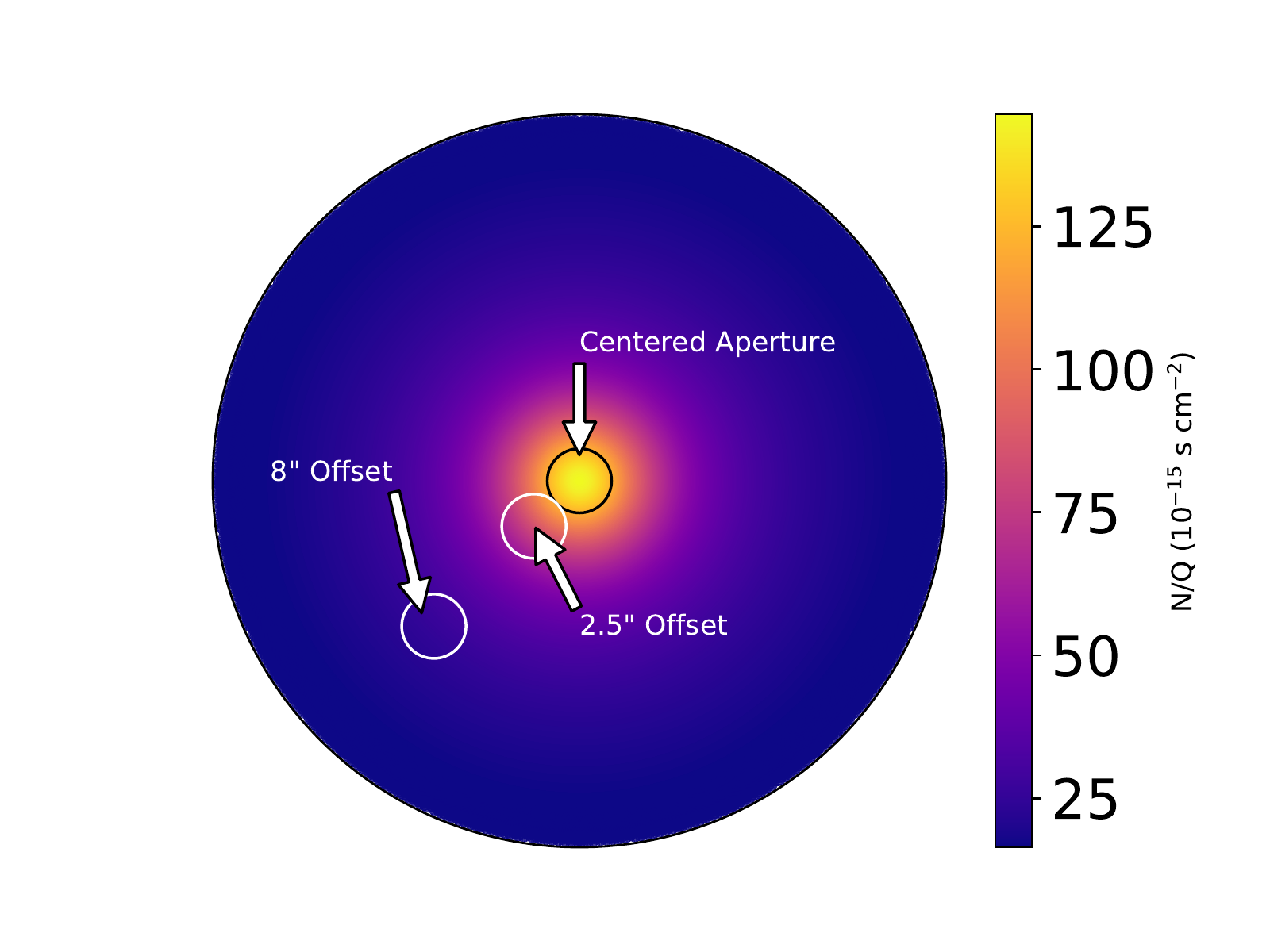}
    \caption{Diagram showing approximate points of the COS aperture for each of the settings described in Table \ref{tab:observations} relative to the coma of 46P. Each white circle represents the field-view-of the COS aperture subtended at $\Delta$=0.22 au and the radius of the full image is 12", approximately 1910 km.  The background image is a simple Haser model for \ce{H2O} at r$_{h}$ = 1.18 au,  which was not observed in the COS bandpass, to provide context for the type of flux decrease that was expected if dissociative electron impact emission were present. The image is independent of production rate and thus has units of s cm$^{-2}$ . The offset direction is not necessarily representative of the as-executed offset angle. }
    \label{fig:cos_aperture_overlay}
\end{figure*}

In this paper we discuss observations of 46P/Wirtanen obtained January 9-20, 2019 with the Cosmic Origins Spectrograph (COS) on the Hubble Space Telescope (HST) between wavelengths of 900 and 1430~\AA. In Section \ref{sec:Obs} we review the observing strategy, as well as the COS instrument and COS data reduction. The reduced spectra, derived column densities, and calculated production rates for assumed parent species are discussed in Section \ref{sec:Results}. Section \ref{sec:Disc} compares these results to previous apparitions of 46P as well as studies from the 2018-2019 apparition. Our conclusions are presented in Section \ref{sec:Summary}.

\section{Observations}\label{sec:Obs}
To access the same region of the FUV probed by the \textit{Rosetta} Alice ultraviolet spectrograph \citep{stern2007alice} that escorted comet 67P/Churyumov-Gerasimenko for over two years we executed observations with the Cosmic Origins Spectrograph on the Hubble Space Telescope with the G130M grating \citep{green_cosmic_2012}. For these FUV observations two separate cross delay line detector arrays are used, each with 16384 spectral and 1024 spatial pixels. Not all pixels are exposed to incoming light. The G130M grating provides a resolving power between 1.75-2.25$\times$10$^{4}$ over the bandpass between 900 to 1430~\AA. The narrow aperture of the COS instrument, 2.5" in diameter, allowed a discrete probing of the inner 1350 kilometers of 46P's coma, as shown in Figure \ref{fig:cos_aperture_overlay}. Two instrument settings were used, one centered at 1096~\AA\ and the other at 1291~\AA\,  for three pointings (centered, 2.5", and 8" offsets) chosen to search for near-nucleus coma emissions and dissociative electron impact like those observed at 67P \citep{feldman_measurements_2015,feldman_nature_2016,bodewits2016changes,chaufray_rosetta_2017,feldman_fuv_2017}, a transition zone near the collisionopause \citep{mandt2016rpc}, and a radius beyond the expected collisional zone where only photodissociation and prompt emission were expected. Spectra taken with a central wavelength of 1096~\AA\ were able to capture wavelength ranges between 900 and 1230~\AA\ in hopes to detect the strongest transitions of the Lyman series of hydrogen atoms in the coma: Lyman-$\alpha$ and -$\beta$, at 1215.72 and 1025.7~\AA, respectively. Properties of each COS exposure from Guest Observer (GO) program 15625 (PI: D. Bodewits) are presented in Table \ref{tab:observations}.

For this set of observations it is important to note the low heliocentric velocity of 46P, varying between 7.70 and 9.9 km s$^{-1}$ (Table \ref{tab:observations}). This low velocity decreases the strength of the Swings effect \citep{1941LicOB..19..131S}, by which the redshift or blueshift of the solar spectrum relative to the comet can enhance or decrease atomic and molecular fluorescence efficiencies, and therefore limits variation of our calculated efficiencies from typical values at 1 au. For the wavelength range covered by the G130M spectral element (900 - 1430~\AA), the feature that benefits the most from this effect is cometary sulfur emission at 1425~\AA, which is produced by resonance scattering of the solar \ion{S}{1} feature by S atoms \citep{roettger_iue_1989,feldman_fuv_2017}, and thus increases the fluorescence efficiency, making it easier to detect. However, the low geocentric velocity makes it difficult to resolve any extended geocoronal emission contribution to the Lyman-$\alpha$ emission at 1215.6~\AA\ \citep{mccoy1992hydrogen}. We discuss our method of geocoronal emission subtraction in Section \ref{Sec:DataReduction } and analysis of the line shape of Lyman-$\alpha$ in Section \ref{sec:Results}. 

Observations were scheduled after the comet's close approach between January 9 and 20 of 2019 to mitigate two factors. The first, and most pressing, was that HST would have difficulty tracking 46P with the accuracy required to keep the inner coma within the narrow aperture at close approach due to the high relative motion across the sky. By scheduling observations for a period when 46P was at a larger geocentric distance the on sky motion of the comet was within the HST maximum slew rate capability. The second issue lies with the characteristics of observing cometary comae; comets appear brightest when their coma are distant and therefore more concentrated on the sky. By observing the inner coma when the comet was just $\sim$0.20 au away, much of the light from the extended coma is excluded, rendering the object relatively faint by typical spectroscopic standards, but significantly brighter than if observations had been conducted at 0.07 au at closest approach. This improved the acquisition likelihood as well as spectroscopic data. 

We note that there are contemporaneous observations taken with the Space Telescope Imaging Spectrograph (STIS) in the same observing campaign (GO-15625 PI: D. Bodewits). These will be discussed in a separate publication (Venkataramani et al, in prep). All observations used in this publication are available in the Mikulksi Archive for Space Telescopes\footnote{\url{https://archive.stsci.edu/}}. 

\begin{longrotatetable}
\begin{deluxetable*}{| c | c | c |c | c | c | c |c | c | c | c}
    \tablecaption{46P/Wirtanen HST COS Observation Log}
    \startdata
     Observation ID & Date &  Center  & Exposure & Offset & $\Delta$\tablenotemark{a,b}  & $\dot{\Delta}$\tablenotemark{c} & Heliocentric & Heliocentric &  Offset Distance\tablenotemark{d} \\
     & (UTC) &  Wavelength (\AA) & Time (s) & Angle (")  & (AU) & (km/s) &  Distance\tablenotemark{a} (AU) & Range-rate\tablenotemark{a} (km/s) & at 46P (km) \\
    \hline
    ldx60901&2019-01-15 17:19:49&1096.0&1630.18&0&0.2&13.15&1.15&8.99&0.0\\ 
    ldx61001&2019-01-16 13:57:17&1096.0&1630.21&0&0.21&13.07&1.15&9.17&0.0\\ 
    ldx61101&2019-01-16 17:07:58&1096.0&1630.18&0&0.21&13.13&1.15&9.19&0.0\\ 
    ldx61201&2019-01-17 13:48:50&1096.0&1630.18&0&0.22&13.9&1.16&9.36&0.0\\ 
    \hline
    ldx60101&2019-01-09 18:12:10&1291.0&1700.16&0&0.17&9.91&1.12&7.7&0.0\\ 
    ldx60201&2019-01-10 21:24:45&1291.0&1660.19&0&0.18&11.64&1.13&7.96&0.0\\
    \hline
    ldx61301&2019-01-18 15:13:47&1096.0&1630.18&2.5&0.22&14.27&1.16&9.57&402.0\\ 
    ldx61401&2019-01-13 11:14:44&1096.0&1630.21&2.5&0.19&11.37&1.14&8.52&345.0\\     
    ldx61501&2019-01-14 11:05:31&1096.0&1630.14&2.5&0.2&11.92&1.14&8.73&355.0\\ 
    ldx61601&2019-01-14 07:54:47&1096.0&1630.18&2.5&0.2&11.85&1.14&8.71&354.0\\ 
    \hline
    ldx60301&2019-01-12 17:46:49&1291.0&1700.13&2.5&0.19&11.39&1.13&8.37&337.0\\ 
    ldx60401&2019-01-16 10:45:23&1291.0&1700.16&2.5&0.21&12.84&1.15&9.14&377.0\\
    \hline
    ldx62901&2019-01-18 07:15:50&1096.0&1630.14&8&0.22&13.77&1.16&9.5&1274.0\\ 
    ldx63001&2019-01-18 16:49:10&1096.0&1630.18&8&0.22&14.31&1.16&9.58&1289.0\\ 
    ldx63101&2019-01-19 18:13:50&1096.0&1630.11&8&0.23&14.58&1.17&9.78&1328.0\\ 
    ldx63201&2019-01-20 08:30:41&1096.0&1630.11&8&0.23&14.49&1.17&9.89&1350.0\\ 
    \hline
    ldx62501&2019-01-13 16:00:16&1291.0&1750.21&8&0.19&11.56&1.14&8.57&1110.0\\ 
    ldx62601&2019-01-15 07:44:32&1291.0&1750.14&8&0.2&12.43&1.15&8.91&1167.0\\ 
    \hline
    \enddata
    \tablenotetext{a}{Queried from JPL Horizons}
    \tablenotetext{b}{Geocentric Distance}
    \tablenotetext{c}{Geocentric Range-rate}
    \tablenotetext{d}{Observations with both identical instrument settings and offsets are co-added to create spectra presented in Figures \ref{fig:spectra_1096} and \ref{fig:spectra_1291} }
    \label{tab:observations}
\end{deluxetable*}
\end{longrotatetable}

\subsection{Data Reduction}\label{Sec:DataReduction }
All COS data were processed with the CALCOS pipeline before any further reduction \citep{Rafelski2018}. The CALCOS pipeline accounts for flatfield correction, pulse height filtering, geometric correction, thermal corrections, and flux calibrations of raw COS spectra. Following the CALCOS reduction each individual spectrum had a scaled airglow spectrum subtracted. Attempts were made to scale this spectrum by average angle to Earth-horizon, but ultimately given the non-detection of \ion{N}{1} 1200~\AA, which was the benchmark for the importance of airglow and geocorona features, the integrated flux between 1198 and 1202~\AA\ was used to scale an averaged airglow spectrum which was then subtracted from each x1dsum spectrum before any coaddition. Airglow spectra were obtained from the HST COS deliberate airglow observation page\footnote{\url{https://www.stsci.edu/hst/instrumentation/cos/calibration/airglow}}.

For emission features with such large signal-to-noise ratios in individual spectra we can extract integrated line brightnesses from those calibrated spectra. To search for emission features that may have emissions too faint to observe in the original COS data we can both bin the spectrum and co-add spectra with the same location relative to the nucleus. Binned and co-added spectra are shown in Figures \ref{fig:spectra_1096} and \ref{fig:spectra_1291}, while the high signal-to-noise co-added but unbinned Lyman-$\alpha$ and \ion{O}{1} 1302 \AA\ triplet emissions are shown in Figures \ref{fig:ly_al_profile}, \ref{fig:ly_al_gaussian}, and \ref{fig:oi_triplet_zoom}.

All spectral fluxes recorded by the CALCOS pipeline are in units of ergs~cm$^{-2}$~\AA$^{-1}$~s$^{-1}$ and are converted to units of Rayleighs, a surface brightness unit useful for extended objects. One Rayleigh corresponds to 10$^{6}$~photons~cm$^{-2}$~s$^{-1}$~sr$^{-1}$, so a conversion function is used to convert the COS spectral data first from an energy flux in ergs to a photon flux based on photon energy, then to a surface brightness. This is done via the equation
\begin{equation}
    B(R) = (\frac{4\pi}{10^{6}}) \Omega F
\end{equation}
where $\Omega$ is the solid angle of the COS aperture in steradians, 1.84$\times$10$^{-9}$ sr, and \textit{F} is the flux from the object in photons cm$^{-2}$~s$^{-1}$~\AA$^{-1}$. 

\section{Results}\label{sec:Results}

\subsection{Spectra}

\begin{figure*} 
    \centering
    \includegraphics[width=0.75\linewidth]{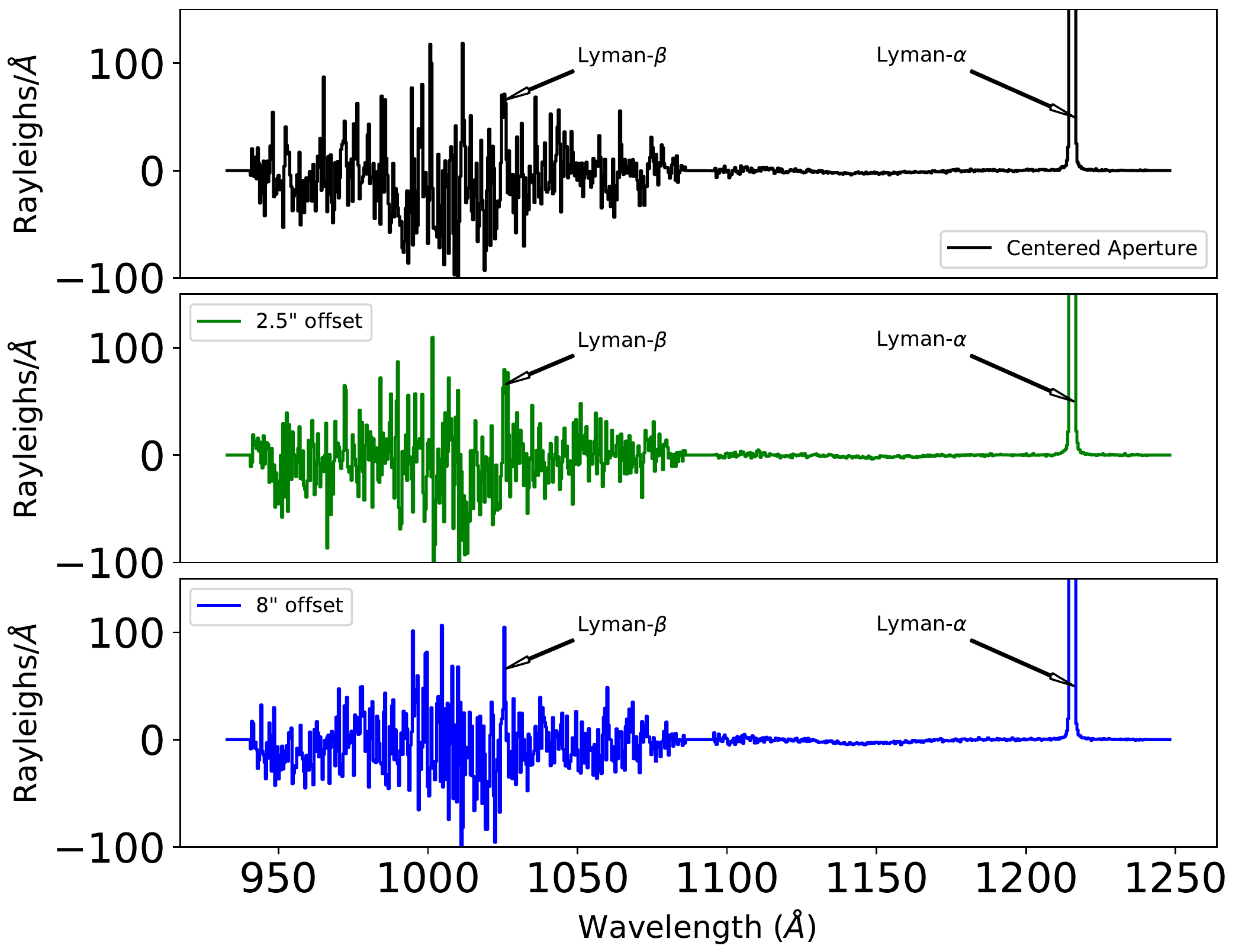}
    \caption{Co-added spectra with 39 pixel binning to improve signal in emission features with low flux, specifically Lyman-$\beta$ near 1026~\AA. Note that only Lyman-$\alpha$ at 1215.7~\AA, is easily identifiable. An enlarged high resolution profile of Lyman-$\alpha$ is presented in Figure \ref{fig:ly_al_profile}. }
    \label{fig:spectra_1096}
\end{figure*}

\begin{figure*}
    \centering
    \includegraphics[width=0.75\linewidth]{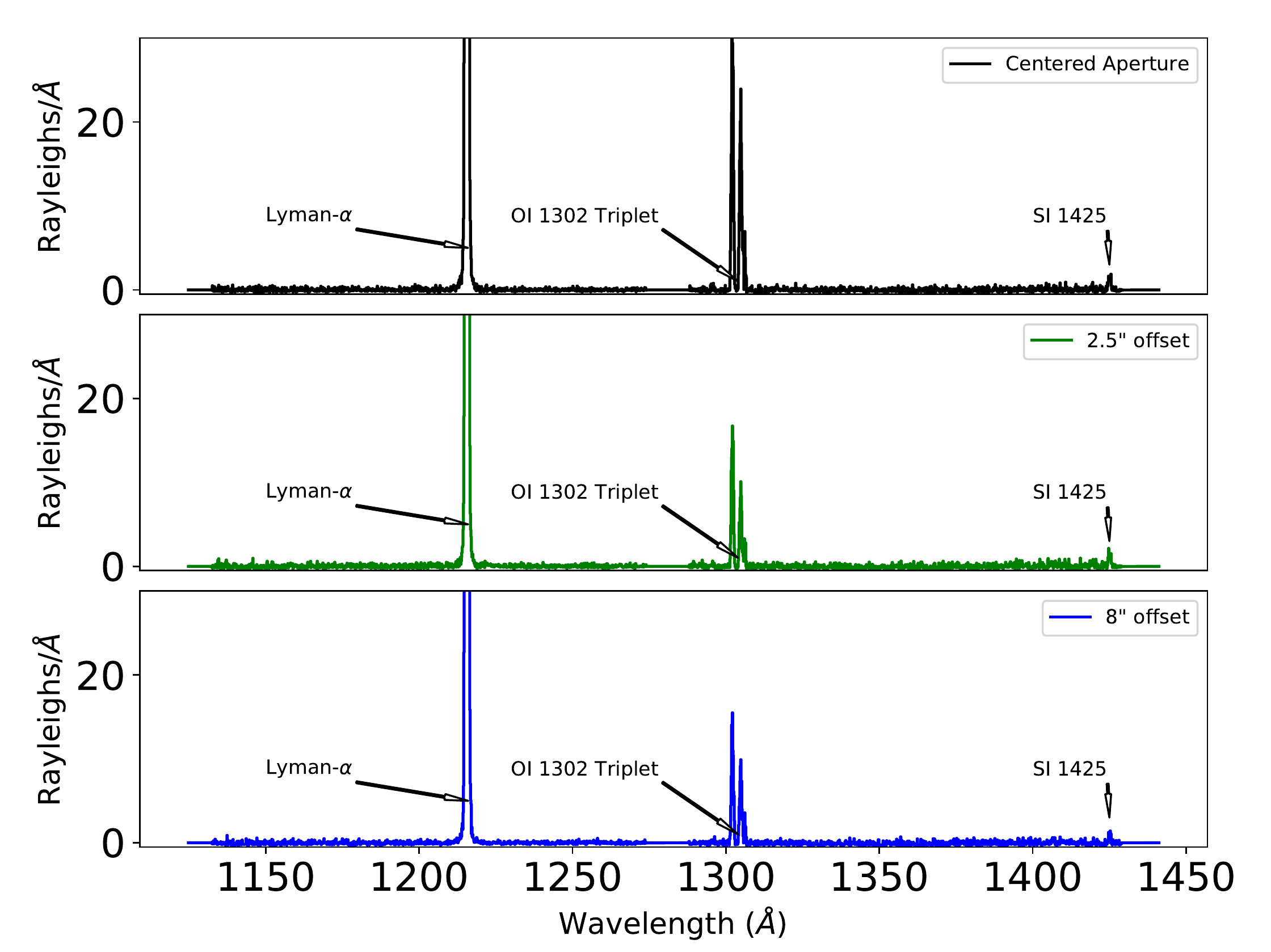} 
    \caption{Co-added spectra with 13 pixel binning to improve signal in emission features with low flux. In addition to Lyman-$\alpha$ emissions at 1215.7~\AA\ the \ion{O}{1} 1302~\AA\ multiplet is easily identified and presented in Figure \ref{fig:oi_triplet_zoom}. With the exception of \ion{S}{1} 1425~\AA, shown enlarged in Figure \ref{fig:sulfur_zoom}, no other significant atomic or molecular emissions were identified. }
    \label{fig:spectra_1291}
\end{figure*}

Co-added spectra for each pointing offset and wavelength setting are presented in Figures \ref{fig:spectra_1096} and \ref{fig:spectra_1291}. We strongly detect the emission features of three atomic species in these data: H (1215.7 and 1025.7 \AA\ ), O ( 1302~\AA\ triplet ) , and S (1425~\AA\ ). 

\begin{figure}
    \includegraphics[width=0.45\textwidth]{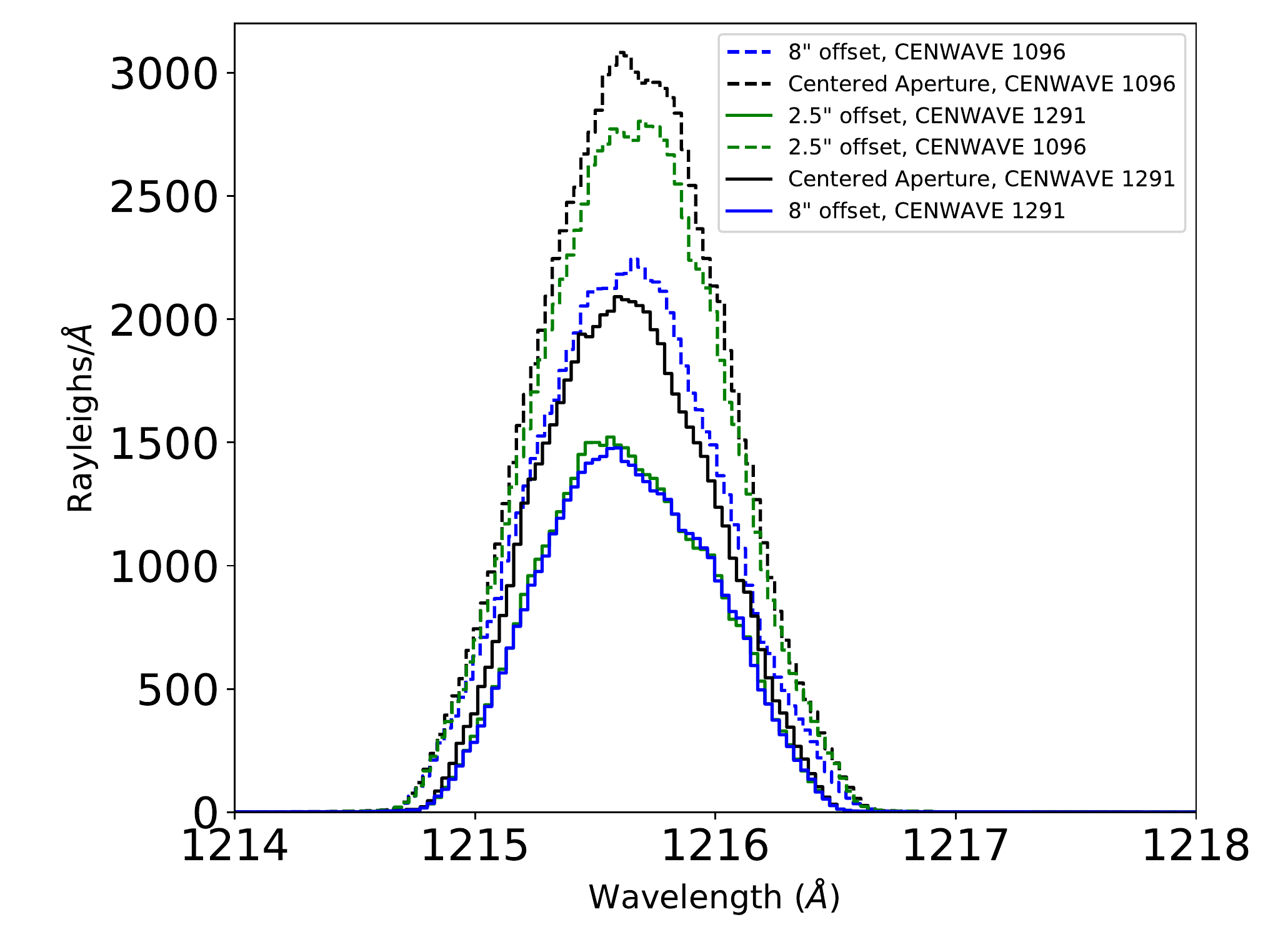}
    \caption{Profile of the Lyman-$\alpha$ emission for each co-added pointing and central wavelength setting with no binning applied. The contribution of geocoronal and interplanetary hydrogen emission to the cometary spectrum is relatively constant for each offset and is minimal, if not negligible.}
    \label{fig:ly_al_profile}
\end{figure}

\begin{figure}
    \includegraphics[width=0.45\textwidth]{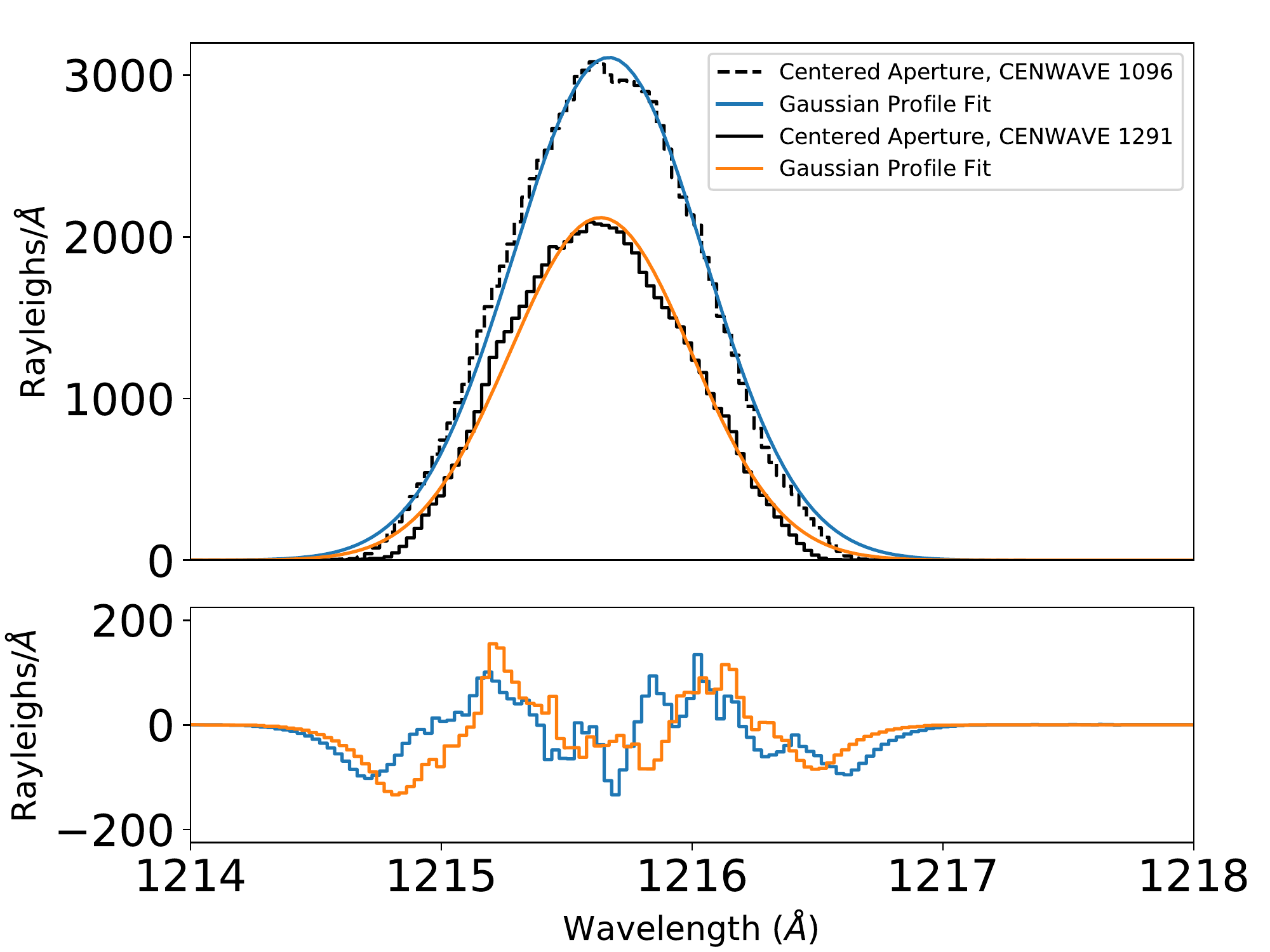}
    \caption{Profile of the Lyman-$\alpha$ emission for the two comet-centered co-added spectra and central wavelength setting with no binning applied. A gaussian fit to the data is plotted and the residuals resulting from subtracting the model from the data are plotted below.}
    \label{fig:ly_al_gaussian}
\end{figure}

 Lyman-$\alpha$ has been used frequently to measure the hydrogen coma from observations with much larger fields of view than HST COS (degrees as opposed to arcseconds) and is a powerful diagnostic for determining the water production rates of comets \citep{mccoy1992hydrogen,combi_iue_1992,combi_hubble_1998,combi2019survey,mayyasi2020lyalpha,combi_comet_2020}. While the Lyman-$\beta$ transition was detected in our coadded spectra, it was at a much lower significance due to increased detector noise blueward of 1100~\AA.  Lyman-$\alpha$ is very strong but proper interpretation of the brightness requires modeling the distribution and radiative transfer of H atoms in the inner coma. The minimal geocoronal contribution is due to the fortunate observing geometry for 46P's apparition, which required nightside observations for HST, and the narrow Field of View (FOV) of the COS instrument. Similarly, the ecliptic longitude and latitude of 46P placed it in the "downwind" interstellar direction, which has a low IPM contribution to the Lyman-$\alpha$ emission of between 300 and 400 Rayleighs \citep{pryor2013lyman}. A high resolution profile of the Lyman-$\alpha$ emission is shown in Figure \ref{fig:ly_al_profile} for each co-added pointing offset, and shows the dominance of cometary hydrogen emissions for the COS observations, with only minor contributions from both the Earth's geocorona and the interplanetary medium in this particular dataset relative to the brightness of the coma. Emission from the 2-1 transition of deuterium at 1215.33~\AA\ is not clearly present when subtracting off a best fit gaussian (Fig. \ref{fig:ly_al_gaussian}). We note that the Ly-$\alpha$ profile of both the airglow and cometary emissions is poorly fit by both Cauchy-Lorentzian and Voigt profiles, and is instead best characterized with a Gaussian profile with a $\sigma$ between 0.36-0.38~\AA, corresponding to a velocity of 83-87.5 km/s. This is much larger than the expected velocities of H atoms in the inner coma (8-26 km/s) \citep{combi2004gas}, and is therefore the limiting resolution of our observations and not indicative of the actual emission profile.The Gaussian fit overpredicts flux at the peak of the emission feature by a small amount and underpredicts flux on the wings but the effect is nearly symmetric, making it difficult to detect deuterium emissions or the IPM contribution directly. We note that this is larger than the expected line spread function of the COS instrument, due to the aperture filling observations of the comet. The 3$\sigma$ resolution of the filled COS aperture is measured to be $\sim$1~\AA\ in our spectra, similar to previous cometary COS observations and airglow measurements. 

\begin{figure}
    \includegraphics[width=0.45\textwidth]{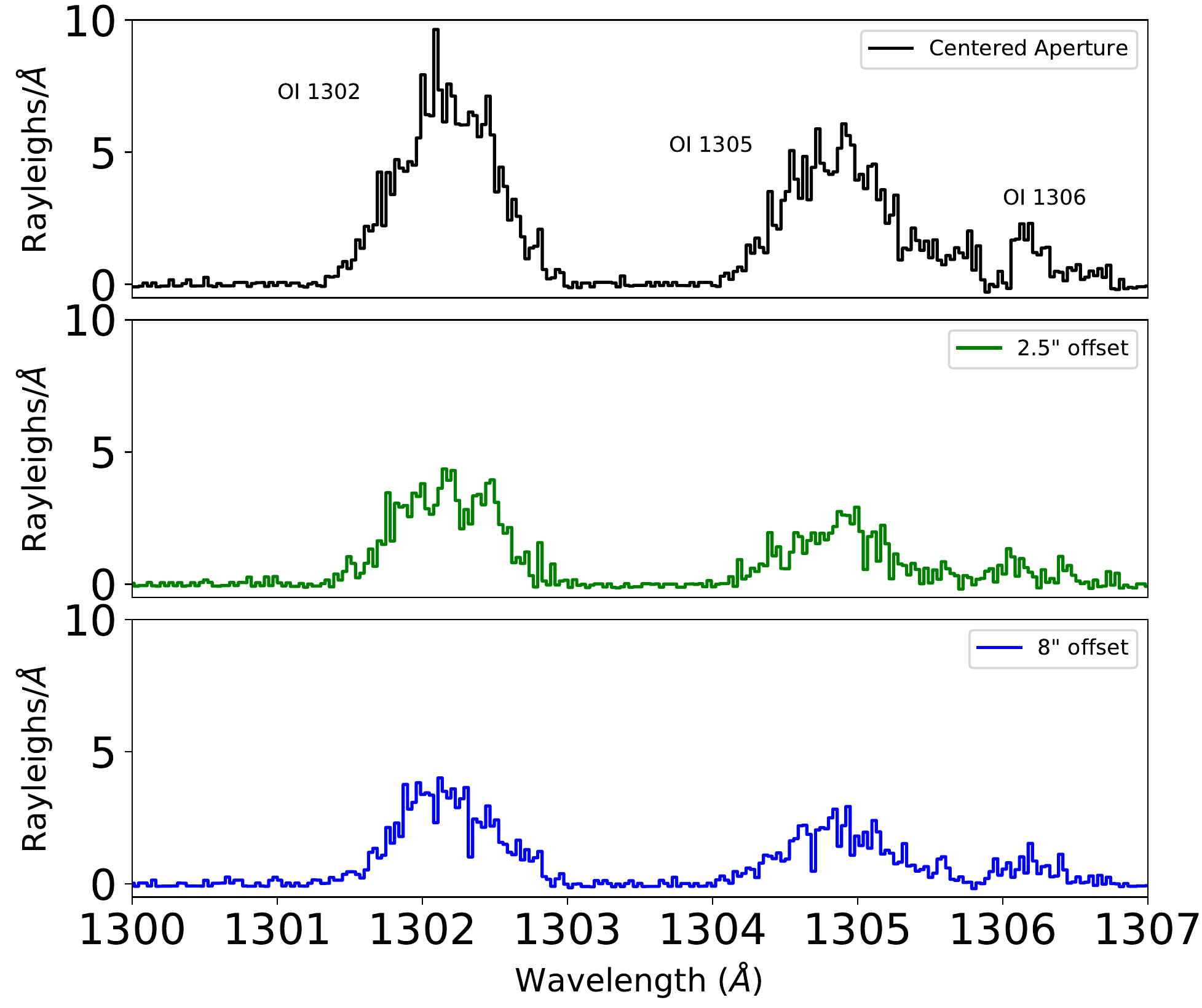}
    \caption{Enlarged section of Figure \ref{fig:spectra_1291} to show \ion{O}{1} triplet emission between 1302 and 1306~\AA. This spectrum has only been binned by a factor of 3 to show line profiles and relative shapes. }
    \label{fig:oi_triplet_zoom}
\end{figure}

The atomic oxygen triplet emission at 1302, 1304, and 1306~\AA\ has an excellent signal-to-noise ratio for all pointings. The spectral resolution provided by the G130M element from COS allows this triplet to be completely resolved (Fig.~\ref{fig:oi_triplet_zoom}). With no clear evidence of self absorption in the \ion{
O}{1} triplet via inversion at the emission peaks \textit{nor} redistribution of flux from the stronger \ion{O}{1} 1302 \AA\~ feature to other features that would indicate significant optical thickness in the atomic oxygen coma, it is reasonable to allow discussion of the diagnostic potential of the \ion{O}{1} triplet in  Section~\ref{sec:Disc}.

\begin{figure}
    \includegraphics[width=0.45\textwidth]{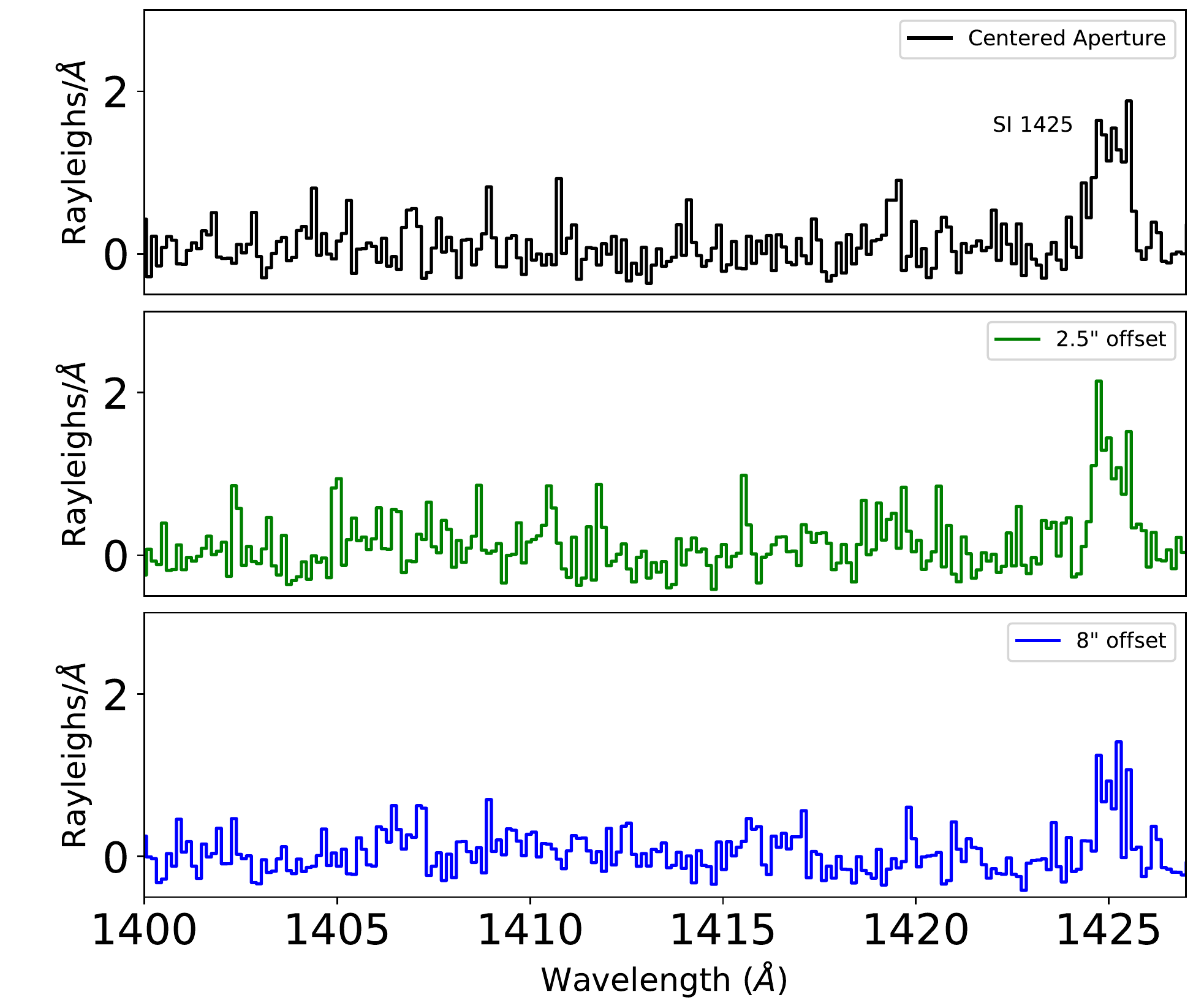}
    \caption{Enlarged section of Figure \ref{fig:spectra_1291} to show \ion{S}{1} 1425~\AA\ emission. }
    \label{fig:sulfur_zoom}
\end{figure}

The last strongly detected atomic emission feature is the triplet of atomic sulfur around 1425~\AA\ in our data, shown in Figure \ref{fig:sulfur_zoom}. As described in Section \ref{sec:Obs} the strength of this feature is enhanced by the low heliocentric velocity of 46P, increasing the fluorescence efficiency of the cometary sulfur from the solar sulfur emission feature.  

\begin{figure}
    \includegraphics[width=0.45\textwidth]{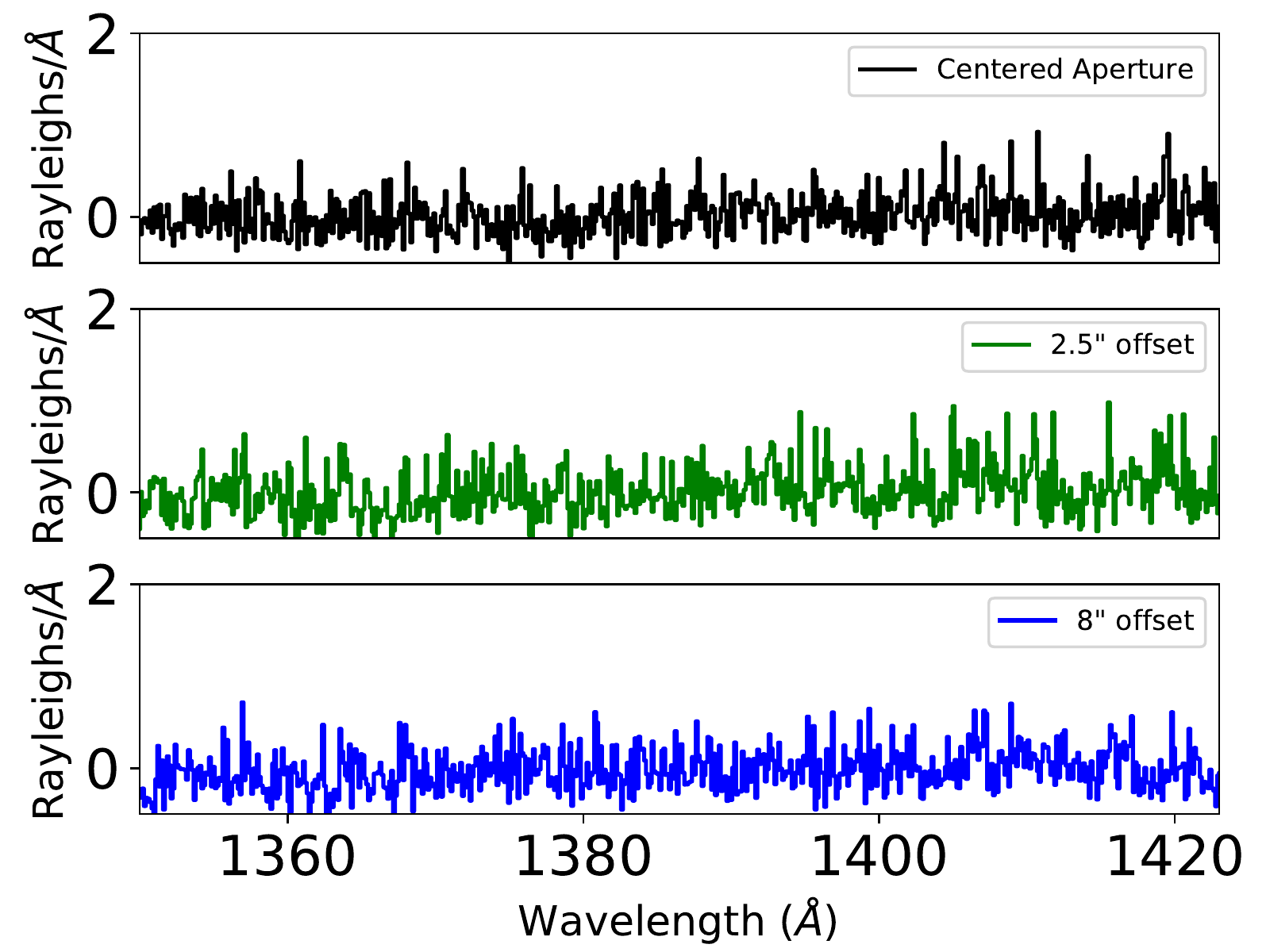}
    \caption{Enlarged section of Figure \ref{fig:spectra_1291} with 13-pixel binning to show lack of CO Fourth Positive Group emissions between 1350 and 1420~\AA\ . \ce{CO} emission features at 1368, 1384, 1392, and 1420~\AA\ are absent.  }
    \label{fig:co_zoom}
\end{figure}

We note that several atomic and molecular features often observed in cometary FUV spectra \citep{weaver_iue_1981,feldman2018far,feldman_fuv_2017} were not detected; we do not see any evidence of the \ion{C}{1} 1277 or \ion{C}{2} 1335~\AA\ emission that could indicate either dissociative electron impact emission of \ce{CO2} or \ce{CO}, or from reflected sunlight from the near-nucleus dust or the nucleus itself. This non-detection is supported by the lack of \ce{CO} Fourth Positive Group emissions between 1350 - 1430~\AA\, which produces a number of features that would be identifiable in our wavelength range (Fig. \ref{fig:co_zoom}) \citep{lupu_fourth_2007,bodewits_carbon_2020}. Another notable absentee in particular is \ion{N}{1} 1200~\AA\ emission, which would have implied significant contribution of airglow emissions to the spectrum.

\subsection{Coma Emission Modeling}\label{subsec:coma_emission_modeling}
For each observed atomic emission feature we calculate the heliocentric velocity dependent fluorescence efficiency, or g-factor, from the Einstein coefficients, oscillator strengths, and energy levels available through the National Institutes of Standards and Technology Atomic Spectra Database\footnote{\url{https://www.nist.gov/pml/atomic-spectra-database}}~\citep{NIST_ASD}. To properly account for the solar emission feature shapes we implement the high-resolution \emph{SUMER} spectrum from \citet{curdt2001sumer}, available through the BASS200 archive\footnote{\url{http://www.bass2000.obspm.fr/solar_spect.php}}. The high resolution spectrum is in relative flux units, normalized to solar emission at 680~\AA\ . The solar spectrum from each day is then determined by multiplying the daily averaged flux from the TIMED-SEE instrument \citep{woods1998timed,woods2000timed}\footnote{Available at  \url{https://www.lasp.colorado.edu/lisird/data/timed_see_ssi_l3/}} for a given day of observations at 680~\AA\ by the \emph{SUMER}-averaged relative spectrum to produce a high resolution ($\sim$0.01~\AA\ ) solar spectrum for calculating g-factors. Solar flux is integrated over a bin $\pm$0.7~\AA\ of the red/blueshifted transition wavelength and the g-factor calculated in accordance with Equation 3 in \citet{2004come.book..425F}. 

Once g-factors have been calculated for the detected atomic emission features, average column densities can be retrieved from the measured brightnesses. This is done via the equation:
\begin{equation}
    \bar{N_a} = \frac{(  10^{6}B_i )}{g_i}
    \label{eq:col_dens}
\end{equation}
where $\bar{N_{a}}$ is the atomic column density in molecules cm$^{-2}$, $B_i$ is the integrated brightness of feature in Rayleighs, and $g_i$ is the g-factor of the atomic transition in phts s$^{-1}$ molecule$^{-1}$. Once a column density has been found for the COS aperture, we can attempt to find the production rates of the likely parent molecules for each atomic species by running the two- or three component Haser model for each observation's timestamp heliocentric and geocentric distance.
\begin{equation}
    Q_n = \frac{\bar{N_a}( \pi \rho^{2})}{H(r_h,\Delta)}
\end{equation} 
where $Q_{n}$ is the neutral production rate, $\rho$ is the radius that the COS aperture subtends at the comet in cm, and $H(r_h,r_\delta)$ is the two- or three-component Haser model result integrated as a function of heliocentric and geocentric distance of 46P for the subtended size of the COS aperture at 46P. By multiplying the average column density by the aperture area the total number of molecules in the field can be found for the numerator. The Haser model is numerically integrated along the line of sight for a range of aperture radii, which are then summed to produce a result with units of s, as it has not been multiplied by a production rate.
Production rates for likely parent molecules were calculated using two- and three-component Haser models \citep{Haser1957,1981A&A....95...69F}. Given the small area subtended by the COS aperture at the distance of 46P/Wirtanen, which is only slightly larger than the collisional radius for the outflowing neutrals ($\sim$300 km, Eq.~5 of \citet{1981A&A....95...69F}), we only used these models for the 0" pointing observations. Lifetimes for \ce{H2O}, \ce{H2S}, \ce{CS2}, and \ce{S2} were taken from \citet{huebner_photoionization_2015}, \citet{meier_atomic_1997}, and \citet{AhearnS2} for solar minimum. The velocity for water was assumed to follow the 0.85$r_{h}^{-2}$ km s$^{-1}$ relation from \citet{combi2004gas}, while the sulfur bearing molecules and atoms were given initial velocities of 0.59$r_{h}^{-2}$ km s$^{-1}$ from \citet{jackson_origin_1986}. We note that for the velocity of OH resulting from dissociation of \ce{H2O} we use the velocity of 1.33 km s$^{-1}$ found by \citet{fink2004effect} specifically for 46P/Wirtanen. Velocities of atomic H were set to 18 km s$^{-1}$, atomic O 1.33 km s$^{-1}$, and atomic S to 1 km s$^{-1}$. 

None of our model runs were able to produce reasonable production rates for parent molecules, often reaching values 10-20$\times$ that of \citet{combi_comet_2020}. A well known issue with the Haser model is its difficulty in reaching high enough column densities near the nucleus \citep{1981A&A....95...69F}, so this is rather unsurprising. Similar attempts were made with a vectorial model for H using the publicly available Web Vectorial Model\footnote{\url{https://www.boulder.swri.edu/wvm/}}, and a similar over estimation of the water production rates found. However, the web vectorial model accurately describes the OH column densities observed in the STIS data, so this issue is likely limited to the narrow COS aperture. For this reason we are unable to accurately produce parent molecule abundances from this dataset at this time, and further work is necessary. Therefore, we will limit our analysis to the atomic column densities and production rates.  

\begin{table*}[t]
    \begin{tabular}{ c | c | c | c | c | c }
     Observation ID & Geocentric Distance (au) & Offset Angle (") &  N$_{H}$ (mol cm$^{-2}$)  &  N$_{O}$ (mol cm$^{-2}$) &  N$_{S}$ (mol cm$^{-2}$) \\
      \hline
    ldx60101&0.1691&0.0&1.62e+13&4.62e+13&1.27e+14 \\ 
    ldx60201&0.1756&0.0&3.74e+13&1.47e+14&2.23e+14 \\ 
    ldx60301&0.1865&2.5&1.94e+13&5.73e+13&1.12e+14 \\ 
    ldx60401&0.2089&2.5&2.04e+13&3.79e+13&2.16e+14 \\ 
    ldx60901&0.2159&0.0&3.61e+13&N/A&N/A \\ 
    ldx61001&0.2106&0.0&3.66e+13&N/A&N/A \\ 
    ldx61101&0.2097&0.0&3.56e+13&N/A&N/A \\ 
    ldx61201&0.2205&8.0&2.94e+13&N/A&N/A \\ 
    ldx61301&0.192&8.0&1.38e+13&2.63e+13&1.39e+13 \\ 
    ldx61401&0.1908&2.5&3.11e+13&N/A&N/A \\ 
    ldx61501&0.196&2.5&3.31e+13&N/A&N/A \\ 
    ldx61601&0.1968&2.5&3.20e+13&N/A&N/A \\ 
    ldx62501&0.202&8.0&1.95e+13&3.84e+13&8.18e+13 \\ 
    ldx62601&0.2226&2.5&3.53e+13&N/A&N/A \\     
    ldx62901&0.2045&0.0&4.10e+13&N/A&N/A \\ 
    ldx63001&0.223&8.0&3.16e+13&N/A&N/A \\ 
    ldx63101&0.2297&8.0&2.72e+13&N/A&N/A \\ 
    ldx63201&0.2336&8.0&2.86e+13&N/A&N/A \\ 
    \hline
    \end{tabular}
    \caption{Derived column densities for H, O, and S in the near-nucleus coma of 46P/Wirtanen from COS data taken between January 8 and 20, 2019. 1-$\sigma$ error bars on the column densities are dominated by variability in the solar flux and are placed at 30\% of each value. Values for N$_{O}$ and N$_{S}$ are listed as N/A when the \ion{O}{1} and \ion{S}{1} emissions were not captured in the CENWAVE 1096 spectra.
    \label{tab:cos_col_densities}}
\end{table*}

\section{Discussion}\label{sec:Disc}
Due to the higher than expected activity of 46P in past apparitions, it may be suggested that 46P could be analogous to 103P/Hartley 2, where \ce{CO2} gas drove out large amounts of water ice grains which resulted in water production rates that exceeded what was expected if water was released by the surface of the comet alone \citep{bertaux_lyman-alpha_1999,fink2004effect,a2011epoxi,kelley2013distribution,protopapa2014water,combi2019survey,lis_terrestrial_2019}. These icy grains are short lived ($\sim$10$^{4}$ s at 1 au) and therefore likely to sublimate throughout the COS aperture for each of the pointings if present in the coma of 46P \citep{yang_comet_2009,yang_multi-wavelength_2014,protopapa_icy_2018}.
The close approach to Earth of 46P/Wirtanen in December 2018 just over two years after the end of the \textit{Rosetta} mission offered a timely opportunity to investigate this question, as well as probe the electron impact plasma environment. Comet 46P/Wirtanen was the primary target of the \textit{Rosetta} mission before a launch delay necessitated changing the target to comet 67P/Churyumov-Gerasimenko. As such, there is a substantial amount of literature dedicated to 46P published between 1996 and 2004 regarding the activity and volatile abundances in previous apparitions. In contrast to 67P, which had a relatively low water production rate for its size (3$\times$4$\times$5 km \citep{sierks2015nucleus}) of 5-7$\times$10$^{27}$ s$^{-1}$ \citep{hansen_evolution_2016}, 46P has been more active than expected for its relatively small size (r=0.63 km\footnote{\footurl} ), \citep{1998A&A...335L..25L,lis_terrestrial_2019} , with peak water production rates between 0.7-5$\times$10$^{28}$ s$^{-1}$ \citep{2003A&A...412..879G,2007MNRAS.376.1399G,2010PASJ...62.1025K,combi_comet_2020}. 
The chemical compositions of the two comae are exceedingly different as well; while the coma of 67P was shown to contain significant levels of \ce{CO2} and \ce{CO} \citep{bockelee-morvan_evolution_2016,feldman_fuv_2017}, neither molecule has been directly detected in 46P's coma despite detections of \ce{CO2}$^{+}$ in UV spectra from 1998 \citep{Stern1998,altwegg_composition_1999} and observations of the same mid-UV bandpass in Venkataramani et al. (in prep). A summary of production rates for common cometary species previously measured in the coma of 46P/Wirtanen is given in Table \ref{tab:prior_prod_rates}. Given the previously known differences between the two comets it's intriguing to find that close examination of our FUV spectra yield even more discrepancies between them. 
\begin{table*}
\centering
    \begin{tabular}{c | c | c | c }
    Species & Production Rate(s) (10$^{27}$ mol s$^{-1}$) & Dates & Refs. \\
    \hline
    \ce{H2O} & 3-50 & 1998,2008 & 1,2,3,4,5 \\ 
    \ce{CN} & 0.00069 - 0.039 & 1998 & 1,2,3,4 \\
    \ce{C2} & 0.0005 - 0.065 & 1998 & 1,2,3,4 \\ 
    \ce{CS2} & 0.02 & 1998 & 5  \\
    \end{tabular}
    \caption{Measured production rates of common volatiles for 46P/Wirtanen during previous apparitions. References are 1) \citet{1998A&A...335L..37F}, 2) \citet{1998A&A...335L..46S}, 3) \citet{1998A&A...335L..50F}, 4) \citet{altwegg_composition_1999}, and 5) \citet{Stern1998}.
\label{tab:prior_prod_rates}}
\end{table*}
The measured spectra of 46P are noticeably devoid of many of the emission features observed by the Alice ultraviolet spectrograph in the inner coma of 67P/Churyumov-Gerasimenko \citep{feldman_nature_2016,feldman_fuv_2017} or those reported in the review of COS comet observations by \citet{feldman2018far}. We focus the discussion here on the Lyman $\alpha$, \ion{O}{1} triplet, and \ion{S}{1} 1425~\AA\ emissions and what they indicate about our models of the near-nucleus emissions as well as the implications for future observations. We note that our observations only overlap with the four comets reported in \citet{feldman2018far} between 1400 and 1430~\AA\ in this particular paper, limiting comparison to atomic sulfur and \ce{CO} abundances.
\subsection{Upper limits on CO Abundance}
The non-detection of the \ce{CO} Fourth Positive group emissions between 1350 and 1420~\AA\ allows us to place an upper limit on both the column density and production rate of carbon monoxide (Fig. \ref{fig:co_zoom}). Using wavelengths, Einstein coefficients, and oscillator strengths for the first seven vibrational transitions from \citet{morton1994compilation} and \citet{beegle1999high} we calculate a total fluorescence efficiency for the (4-0) band of 1.44$\times$10$^{-7}$ phts s$^{-1}$ for the average heliocentric distances and velocities for the comet-centered pointings (Table \ref{tab:observations}). Given observational sensitivities down to 1 Rayleigh for the integrated band wavelength range between 1419 and 1421~\AA, we can then calculate an upper limit for the average column density of \ce{CO} within the COS aperture using Equation \ref{eq:col_dens}. We find that our observations were sensitive to $\bar{N}_{\ce{CO}}$ of 6.9$\times$10$^{12}$ cm$^{2}$. Due to the high spatial resolution of the observations we follow the same model for production of \ce{CO} as \citet{weaver_carbon_2011}. This model argues that the long lifetime of \ce{CO} at 1 AU of 1.5$\times$10$^{6}$ s can be ignored and the production rate derived from $Q=N_{CO}vd$, where $v$ is the velocity of \ce{CO} molecules and $d$ is the diameter of the COS aperture in cm at 46P. Using the same $v$ of 7.8$\times$10$^{4}$ cm s$^{-1}$ as \citet{weaver_carbon_2011} and a $d$ of 3.37$\times$10$^{7}$ cm we place an upper limit on the total production rate of \ce{CO} at 1.8$\times$10$^{25}$ mol s$^{-1}$. This upper limit is near the production rate measured for 103P/Hartley 2 by \citet{weaver_carbon_2011}.  Taking into account the 30\% error on these g-factors, this would put the 3-$\sigma$ upper limit for the \ce{CO}/\ce{H2O} ratio for the period of our observations at  $<$8$\times$10$^{-2}$.  This upper limit would place 46P/Wirtanen in the middle of the range of CO/\ce{H2O} values in \citet{biver_complex_2019}.

\subsection{Upper limits on dissociative electron impact emission}
The lack of any \ion{O}{1}~ 1356~\AA\ is a robust indicator that dissociative electron impact is not a significant source of emissions for the inner coma of 46P. Sampling the noise in the co-added spectra near the expected \ion{O}{1} 1356~\AA\ feature shows that we were sensitive to approximately 1 Rayleigh integrated over the 1352-1358~\AA\ region. Our calculations for the emission rate of \ce{O2} and \ce{H2O}, from \citet{kanik_electron_2003} and \citet{makarov_kinetic_2004}, respectively, show that for an expected Maxwellian electron distribution characterized by a temperature of 25 eV and an electron density of 50 electrons cm$^{-3}$ the COS co-added spectrum brightness upper limit translates to a column density of $\sim$4$\times$10$^{14}$ cm$^{-2}$ for \ce{O2} and $\sim$4$\times$10$^{16}$ cm$^{-2}$ for \ce{H2O}. If dissociative electron impact occurs in the near nucleus coma the total affected column of either molecule must be less than either of these stated values. We note that this upper limit for the \ce{O2} column density is of similar magnitude to the column densities of \ce{O2} detected by the Alice ultraviolet spectrograph onboard the \emph{Rosetta} mission, but was acquired over a much larger FOV subtended at the comet (Noonan et al, in prep.). With observations sensitive down to 0.1 R/\AA, this means that the portion of the inner coma of 46P susceptible to large scale dissociative electron impact (d$_{c}$ $<$ $\sim$ 50 km) could not have had a total integrated brightness greater than 1 R. This strongly implies that dissociative electron impact in the inner coma is unobservable from HST even with an extremely favorable apparition. The relevant scale is simply too small to be captured adequately, even with a 2.5" FOV and a low geocentric distance.

\subsection{Atomic emission}
Given the robust detections of three atomic emission features we investigate the properties available from each. 
\subsubsection{Lyman-$\alpha$ emissions}
Emission from the 2-1 transition of the hydrogen atom is easy to detect but often difficult to analyze, especially on small spatial scales. Line profiles of the Lyman-$\alpha$ emission from each of the co-added spectra show deviation from standard Voigt profiles, indicating the presence of effects from optically thick column densities near the nucleus convolved with the line spread function of the COS instrument at 1215~\AA. We do not detect the deuterium 2-1 electron transition (rest wavelength 1215.33~\AA) due to the $\sim$1~\AA\ resolution of the COS data from which a direct D/H ratio could be calculated. However, we can place an upper limit on the abundance of deuterium by subtracting a best fit gaussian profile from the Lyman-$\alpha$ emission and summing the remaining emission between 1215.0 and 1215.7~\AA\ to find integrated brightnesses between 68 and 330 R for the co-added spectra. Given the similar g-factors for the D (1-0) transition and the H (1-0) transition, this corresponds to an upper limit on deuterium column density abundance between 0.29 and 1.43$\times$10$^{11}$ cm$^{2}$, for a conservative upper limit on the D/H ratio of 46P/Wirtanen of 0.005, approximately an order of magnitude larger than is typical for Jupiter family comets and measured for 46P/Wirtanen \citep{Altwegg1261952,lis_terrestrial_2019}. A more involved effort to model the emission profile of Lyman-$\alpha$ emission will be attempted in future work.

Deriving a water production rate from the Lyman $\beta$ emission is useful to compare with the near daily water production rates calculated by \citet{combi_comet_2020} for the 2018-2019 apparition of 46P/Wirtanen. Finding integrated brightnesses of the Lyman-$\beta$ feature in coadded spectra, between 65 and 170  Rayleighs, and the associated calculated g-factor between 5.0-5.33$\times$10$^{-6}$ s$^{-1}$, allows us to calculate aperture averaged hydrogen column densities between 1.2 and 3.2$\times$10$^{13}$ cm$^{2}$. These column densities are in agreement with those derived from Lyman-$\alpha$, but our two- and three-component Haser models have difficulty matching these column densities.  \citet{combi_comet_2020} use the Solor Wind ANisotropies (SWAN) instrument and a more involved physical model of the H atom distribution, in addition to large coma images within 8 degrees of the nucleus; here we are specifically focusing on the coma within 2.5" of the nucleus, a radius that is approximately 23,000 times smaller. We recognize that the two-component Haser model will not accurately represent the environment within the $\sim$340 km diameter aperture of COS at 46P/Wirtanen, and a simple model was unable to produce production rates within the uncertainty range of the values reported in \citet{combi_comet_2020}. This discrepancy discourages us from using the Haser model for the offset pointings; a hybrid model is required to properly analyze both the emission feature profile and spatial profiles and will be described in a future publication.

\subsubsection{\ion{O}{1} 1302~\AA\ triplet emissions}

\begin{figure}
    \centering
    \includegraphics[width=\linewidth]{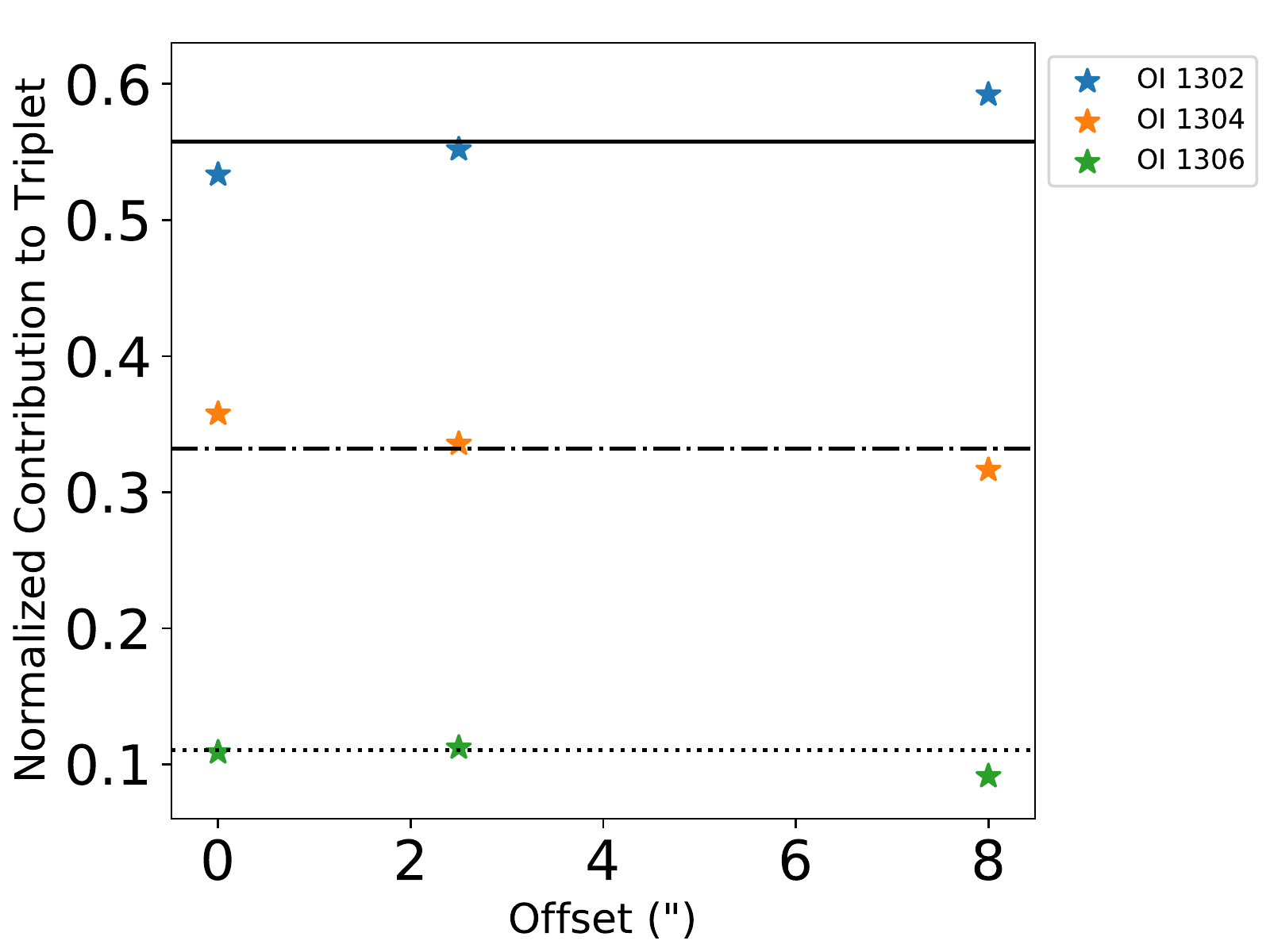}
    \caption{Normalized ratio of \ion{O}{1} triplet emission at 1302, 1304, and 1306~\AA\ for co-added spectra at each pointing offset in arcseconds. In general the O $^{3}S^{0}$ J=2 to the O $^{3}$P J=1 transition at 1302~\AA\ becomes more populated as the offset increases, but the line ratios are never in good agreement with the ratios defined by the known $A$ values~\citep{NIST_ASD}. The expected ratios for $A_{I}$/$\Sigma A_{n}$ are shown with a solid black line for \ion{O}{1} 1302~\AA, a dot-dash line for \ion{O}{1} 1305~\AA, and a dotted line for \ion{O}{1} 1306~\AA. Errorbars are smaller than the markers.}
    \label{fig:oi_partition}
\end{figure}

The \ion{O}{1} resonance triplet is resolved here for the first time in a cometary coma, offering insight into the interplay between collisional- and photo-excitation of fragment species. As illustrated in Fig.~\ref{fig:oi_triplet_zoom} each individual transition from $^{3}S^{0}$ to the $^{3}P_{J=2,1,0}$ states is resolved at wavelengths of 1302.2, 1304.9, and 1306.0~\AA. One expects the contribution of each transition to the total triplet emission (e.g. $\frac{F_{1302}}{F_{1302}+F_{1304}+F_{1306}}$) to follow the ratios between known Einstein $A$-values of the three transitions. However, this does not seem to be the case for the \ion{O}{1}~$(^3S_1)$~triplet emission observed in the inner coma of 46P. For co-added spectra at each of the three pointing angle offsets the normalized ratio of each transition feature was determined and is shown in Figure \ref{fig:oi_partition}.  Both the contribution of the $^3S_1$ to $^3P_2$ ($1302$ \AA) and $^3S_1$ to $^3P_0$ ($1306$ \AA) transitions increase at offset 2.5'', with the $1302$ \AA~line further increasing at 8'' offset. Interestingly, the largest deviation from the expected triplet contribution is for the $1302$ \AA~transition to ground. Both $1304$ and $1306$~\AA~transitions decay to the metastable $^3P_1$ and $^3P_0$ levels, with the largest deviations seen between the 0'' and 8'' offsets. All three transitions have $A$ values in order $10^{7}-10^{8}$~s$^{-1}$~\citep{NIST_ASD}, leading to a lifetime around 1.6~ns, indicating that travel and de-excitation of \ion{O}{1} outside the FOV contributes negligibly to the observed ratios. Additionally, uncertainty in the $A$ values cannot be the cause; the $A$ values for the \ion{O}{1} triplet lines are known to within $\sim$3\%~\citep{NIST_ASD}. In all cases, the uncertainty in the intensity changes is considerably less than the observed change in contribution.

We investigated the possibility that the \ion{O}{1} triplet features may contain contributions from the increasing density of S atoms, which have electronic transitions from the $^{3}P^{o}$ to the $^{3}P$ state and have a series of emissions present between 1302 and 1308~\AA. We find the g-factor for the strongest \ion{S}{1} transitions between 1302 and 1305~\AA~to be $\sim$1.1$\times$10$^{-8}$ s$^{-1}$ for the dates in question, and using the atomic sulfur column density calculated from the \ion{S}{1} 1425~\AA\ emission of $\sim$1.7$\times$10$^{13}$ cm$^{-2}$ we find that less than 0.1 Rayleighs can be attributed to \ion{S}{1} emission in the 1302-1306~\AA\ region. 

We note that the line shape of the \ion{O}{1} triplets is well fit by Gaussian rather than Voigt profiles, with $\sigma$ = 0.33~-~0.35~\AA, in the co-added spectra in Figure \ref{fig:oi_triplet_zoom}. The goodness of fit with Gaussian profiles is not unexpected given that the O coma is an extended source. There is no clear evidence that would indicate that the \ion{O}{1} transitions are optically thick.

Assuming a water production rate of $\sim7\times10^{27}$~mol.~s$^{-1}$, the density of molecules in our FOV (100s~km from the nucleus) is in order $7\times10^{17}$~m$^{-3}$. Assuming a collisional cross section in order $10^{-15}$~cm$^{-2}$, the mean free path $\lambda_{\textrm{MFP}} = 1/\sigma\rho$ between collisions is $\sim$3~mm. The collisional frequency, $\nu = v_{\textrm{rms}}/\lambda_{\textrm{MFP}}$, is then found from the root-mean-square velocity of the gas and the local mean free path. Assuming an \ion{O}{1} gas temperature of 100~K yields $v_{\textrm{rms}}\sim$~220~m/s, from which the collisional rate in the near coma follows as $\sim7.6\times10^4$~s$^{-1}$. Thus, following population of the $^3S_0$ state, the collisional frequency is too small by 5 - 6 orders of magnitude to begin contributing via collisional de-excitation, indicating that collisional effects are insufficient for explaining the triplet ratios.

An alternative explanation for the triplet emission contributions may be the incident solar radiation, with stimulated emission enhancing the observed line intensities. The assumed incident solar flux from SUMER (see Sec.~\ref{subsec:coma_emission_modeling}) at the \ion{O}{1} triplet wavelengths is approximately 1:1:1 between the 3 triplet lines, which suggests negligible contributions of heterogenous stimulated emission from $^3S_1$. However, there is extensive literature on understanding O~I triplet emission in the solar spectrum, including detailed radiative transfer models of O~I resonance line excitation~\citep{Bhatia1995} and \citep{Carlsson1993}~(who also found non-Voigt line profiles), polarization~\citep{Anusha2014}, and frequency cross redistribution effects~\citep{MillerRicci2002}. Our co-added spectra were observed over a period of 9 days, during which it may be possible that changes in the solar spectrum preferentially enhanced the $1302$~\AA~emission, though this proposition conflicts with the SUMER solar spectra. Given the negligible contribution from collisional effects, the most likely cause of the deviations in relative triplet emission intensities (Fig.~\ref{fig:oi_partition}) is the incident solar flux. As these lines are also sensitive diagnostics at the \textit{source} of the \ion{O}{1} solar flux, one can expect similar diagnostic potential when observed in an O-rich environment such as cometary comae. In particular, these lines may offer an additional way to distinguish between photon- and electron-dominated environments in future observations.

Understanding atomic oxygen emissions and its implications for high resolution spectroscopy of the near-nucleus coma are a critical component for improving the scientific return from comet spectroscopy. Given the observed differences in \ion{O}{1} triplet emission with offset angle, these transitions may prove to be useful diagnostics of the conditions in the inner coma. At present, developing a time-dependent atomic model of the incident solar radiation producing $^3S_1 \rightarrow ^3P_{J=0,1,2}$ emission is beyond the scope of this work. Further studies of the \ion{O}{1} transitions in the near- and extended-comae of comets is required to fully understand the diagnostic potential of these lines.
\subsubsection{\ion{S}{1} 1425~\AA\ emissions}
The detection of \ion{S}{1} 1425~\AA\ emission was unexpected given the narrow FOV (Figure \ref{fig:sulfur_zoom}).  Such a detection was possible because of the relatively low heliocentric velocity of 46P/Wirtanen during the observations, enabling efficient resonance scattering of the solar \ion{S}{1} feature \citep{roettger_iue_1989,feldman2018far}. Our derived column densities for atomic sulfur in 46P in Table \ref{tab:cos_col_densities} are similar to those derived by \citet{feldman2018far} for both C/2014 Q2 (Lovejoy) and 153P/Ikeya-Zhang, which were measured to have N$_{S}$ of 1.2 and 2.0$\times$10$^{14}$ cm$^{-2}$, respectively. Given the high abundance of S in the inner nucleus it is necessary to identify potential parent molecules from the nucleus. 

A favorite parent molecule for cometary atomic sulfur is \ce{S2}, due to both its short lifetime and single atomic components \citep{AhearnS2,meier_atomic_1997}.  For production rates on the order of 10$^{25}$ mol s$^{-1}$ the \ce{S2} band emission between 2800-3100~\AA\ should be easily detectable \citep{AhearnS2}.  Therefore, the lack of \ce{S2} band emission in STIS observations taken as part of the same campaign (Venkataramani et al. 2020, in prep) suggests that consideration of \ce{CS2} is a reasonable source of \ion{S}{1} 1425~\AA\ emission, via photodissociation into CS and S, is necessary. \ce{CS2} has a relatively short lifetime of 590 s \citep{jackson_origin_1986} compared to the 82000 s lifetime of \ce{H2O} \citep{combi2004gas}, and is just approximately 150 s longer than that of \ce{S2} \citep{meier_atomic_1997}.  However, the previously measured \ce{CS2} production rates between 2-5$\times10^{-2}$ that of \ce{H2O} \citep{Stern1998}, and a similar measurement of CS production rates 1-2$\times10^{-2}$ that of \ce{H2O} from the concurrent STIS observations (Venkataramani et al., in prep) are insufficient to explain the observed abundance of atomic sulfur in the observations. Other molecules identified in \citet{feldman2018far} like \ce{H2S} and \ce{SO2} could also contribute to the atomic sulfur column density, but require two dissociations to produce a sulfur atom. This makes them rather unfavorable as dominant sources for the inner coma, but necessary to consider in future modeling. As stated in other sections it is difficult to derive reasonable production rates of the likely parent molecules from these extreme inner coma column densities with empirical Haser and physical vectorial models, and a more robust modeling approach is required.

\subsubsection{Atomic Production Rates of the near-nucleus coma}
With detections of H, O, and S  we can place some constraints on the atomic production rates of the inner coma of 46P from UV observations. However, we note that these values are representative of the near-nucleus coma, not of the overall cometary abundance, and are therefore difficult to directly compare with other comets. For this reason, we derive the atomic production rates for the near-nucleus coma from the comet-centered observations. From the column densities in Table \ref{tab:cos_col_densities} we derive Q$_{H}$, Q$_{O}$, and Q$_{S}$ with a simple Haser model for column densities, with lifetimes of 1.5$\times$10$^{6}$, 1.8$\times$10$^{6}$, and 9.1$\times$10$^{6}$ s for H, O, and S \citep{huebner_photoionization_2015,meier_atomic_1997}. We use a velocity of 18 km/s for H and for O and S a velocity more reflective of the OH velocity for 46P/Wirtanen as found by \citet{fink2004effect}, 1.33 km/s. We can then use the equation for column density from a simple Haser model:
\begin{equation}
    Q = \bar{N} 2\pi \rho v e^{\rho \beta}
\end{equation}
where  N is the average columnn density across the aperture, $\rho$ is the distance from the nucleus, and $\beta =(v\tau)^{-1}$, where $v$ is the velocity of the atoms and $\tau$ is the lifetime. 

We find that the average atomic production rates for H and O are 3.4$\pm$0.3$\times$10$^{27}$ and 1.3$\pm$0.7$\times$10$^{27}$, respectively. The similarity between Q$_{H}$ and the water production rates of \citet{combi_comet_2020} suggest that the dominant source of H in the inner coma is indeed the first dissociation of \ce{H2O}, and that both Haser and vectorial models are unable to accurately represent the densities observed. The production rate of O is approximately two times lower than that of H. Given that two dissociations of H$_2$O are needed to produce a single O atom the discrepancy between Q$_{H}$ and Q$_{O}$ is not unexpected. \ce{H2O} is therefore capable of producing all observed O and no other abundant oxygen-bearing molecules are needed to explain the observed column densities. This is in agreement with our non-detection of \ce{CO} and \ce{CO2}$^{+}$ in Venkataramani et al., in prep. 

The production rate of sulfur is less easily explained. The inner coma of 46P/Wirtanen has Q$_{S}$ of 2.3$\pm$0.5$\times$10$^{27}$ s$^{-1}$, rivaling the production of hydrogen atoms. As described in the previous section this number is difficult to obtain with the known production rates of sulfur-bearing \ce{CS2}, and the preferred parent \ce{S2} does not have the corroborating emission in the 2600-2900~\AA\ range in STIS data for the needed production rates. \citet{calmonte2016sulphur} provide a variety of sulfur-bearing molecules detected in the near-nucleus environment of 67P/Churyumov-Gerasimenko, but to match the COS observations there are two clear constraints. Whatever the unknown sulfur parent, it must have a lifetime on the order of 100's of seconds and be produced directly either from the nucleus itself or from sublimating grains within the first 10's of km from the nucleus. In addition, the non-sulfur daughter products of the dissociation cannot contribute substantially to either the H or O columns. This is similar to one particular finding of \citet{calmonte2016sulphur}; 27\% of atomic S in the inner coma of 67P/Churymov-Gerasimenko could not be linked to a parent molecule. The Alice UVS also reported significant amounts of atomic sulfur in the coma, with no clearly identifiable parent molecule \citep{2015DPS....4750306F,feldman_fuv_2017}. A similar situation may be present in the inner coma of 46P/Wirtanen, where a large component of S atoms has no clearly identifiable parent molecule, and may be sublimating off of the cometary surface or from grains in the inner coma. This conclusion also warrants a closer look at the production of sulfur in the inner coma of other comets observed with COS, especially C/2014 Q2 (Lovejoy) and 153P/Ikeya-Zhang \citep{feldman2018far}.

\section{Summary}\label{sec:Summary}
In this paper we have presented spectra of 46P/Wirtanen from 900 - 1430~\AA\ taken when the comet was between 0.16 and 0.23 au from Earth and 1.12 and 1.17 au from the Sun. During this period the 2.5" diameter aperture of HST COS subtends between 337 and 402 km at the comet, allowing portions of the near-nucleus coma to be observed. Our results can be summarized as follows:
\begin{enumerate}
    \item We found no evidence of \ce{CO} Fourth Positive Group emission between 1350 and 1430~\AA, and use the (4-0) band at 1420~\AA\ to place a 3$\sigma$ upper limit on the production rate of CO at 3.6$\times$10$^{25}$ mol s$^{-1}$, approximately 8\% that measured for \ce{H2O}.
    \item No evidence of dissociative electron impact was detected via the semi-forbidden \ion{O}{1} 1356~\AA\ emission feature. We place upper limits on the aperture averaged column densities of \ce{H2O} and \ce{O2}   susceptible to dissociative electron impact at $\sim$4$\times$10$^{14}$ cm$^{-2}$ for \ce{O2} and $\sim$4$\times$10$^{16}$ cm$^{-2}$ for \ce{H2O}. 
    \item The \ion{O}{1} 1302~-~1306~\AA\ triplet was resolved for the first time in a cometary coma, yielding relative line ratios which change with offset and are inconsistent with known $A$ values. Future observations will be required to uncover the diagnostic potential of these lines.
    \item Derived atomic production rates of H and O imply that the only substantial source of H and O in the coma is \ce{H2O}. This suggests that \ce{CO2} and \ce{O2} are not abundant in the coma of 46P/Wirtanen for the 2018-2019 apparition. 
    \item The derived production rates of atomic sulfur are only slightly less than that of H. This production rate is difficult to explain with the known sulfur-bearing molecules on 46P and suggests that atomic sulfur may be entering the near-nucleus coma directly from the nucleus or grains very near the surface, similar to 67P/Churyumov-Gerasimenko \citep{calmonte2016sulphur}.
\end{enumerate}
Attempts to derive parent molecule production rates using Haser and vectorial modeling were unable to produce values within reasonable agreement of water production rates from \citet{combi_comet_2020}. Monte Carlo modeling of the inner coma is required to properly interpret observations taken at offsets and perform full analysis of the information contained within the Lyman-$\alpha$ and \ion{O}{1} 1302~\AA\ emission profiles. Given the continued observed abundance of sulfur in the inner nucleus of comets \citep{feldman2018far} we recommend a re-examination of past comets observed with COS and STIS with improved modeling as well as further study of possible pathways for atomic sulfur to be introduced into the inner coma. 

\section*{Acknowledgements}
Based on observations with the NASA/ESA/CSA Hubble Space Telescope obtained at the Space Telescope Science Institute, which is operated by the Association of Universities for Research in Astronomy. All authors extend their sincere thanks to Alison Vick, Tom Brown, Tony Sohn, and William Fischer for helping schedule and execute these challenging observations. Incorporated, under NASA contract NAS5-26555. All authors acknowledge support by HST program number GO-15625 (PI D. Bodewits), which was provided through a grant from the STScI under NASA contract NAS5-26555. Part of this research was conducted at the Jet Propulsion Laboratory, California Institute of Technology, under a contract with NASA.

\bibliographystyle{aasjournal}

\bibliography{46PWirtanen_ads,alice,references_long}

\begin{thebibliography}{}
\expandafter\ifx\csname natexlab\endcsname\relax\def\natexlab#1{#1}\fi
\providecommand{\url}[1]{\href{#1}{#1}}
\providecommand{\dodoi}[1]{doi:~\href{http://doi.org/#1}{\nolinkurl{#1}}}
\providecommand{\doeprint}[1]{\href{http://ascl.net/#1}{\nolinkurl{http://ascl.net/#1}}}
\providecommand{\doarXiv}[1]{\href{https://arxiv.org/abs/#1}{\nolinkurl{https://arxiv.org/abs/#1}}}

\bibitem[{A'Hearn(2011)}]{ahearn_comets_2011}
A'Hearn, M.~F. 2011, Annual Review of Astronomy and Astrophysics, 49, 281,
  \dodoi{10.1146/annurev-astro-081710-102506}

\bibitem[{{A'Hearn} {et~al.}(1983){A'Hearn}, {Schleicher}, \&
  {Feldman}}]{AhearnS2}
{A'Hearn}, M.~F., {Schleicher}, D.~G., \& {Feldman}, P.~D. 1983, \apjl, 274,
  L99, \dodoi{10.1086/184158}

\bibitem[{Altwegg {et~al.}(1999)Altwegg, Ehrenfreund, Geiss, \&
  Huebner}]{altwegg_composition_1999}
Altwegg, K., Ehrenfreund, P., Geiss, J., \& Huebner, W.~F., eds. 1999,
  Composition and {Origin} of {Cometary} {Materials} (Dordrecht: Springer
  Netherlands), \dodoi{10.1007/978-94-011-4211-3}

\bibitem[{Altwegg {et~al.}(2015)Altwegg, Balsiger, Bar-Nun, Berthelier, Bieler,
  Bochsler, Briois, Calmonte, Combi, De~Keyser, Eberhardt, Fiethe, Fuselier,
  Gasc, Gombosi, Hansen, H{\"a}ssig, J{\"a}ckel, Kopp, Korth, LeRoy, Mall,
  Marty, Mousis, Neefs, Owen, R{\`e}me, Rubin, S{\'e}mon, Tzou, Waite, \&
  Wurz}]{Altwegg1261952}
Altwegg, K., Balsiger, H., Bar-Nun, A., {et~al.} 2015, Science, 347,
  \dodoi{10.1126/science.1261952}

\bibitem[{Anusha {et~al.}(2014)Anusha, Nagendra, \& Uitenbroek}]{Anusha2014}
Anusha, L.~S., Nagendra, K.~N., \& Uitenbroek, H. 2014, The Astrophysical
  Journal, 794, 17, \dodoi{10.1088/0004-637x/794/1/17}

\bibitem[{A’Hearn {et~al.}(2011)A’Hearn, Belton, Delamere, Feaga, Hampton,
  Kissel, Klaasen, McFadden, Meech, Melosh, {et~al.}}]{a2011epoxi}
A’Hearn, M.~F., Belton, M.~J., Delamere, W.~A., {et~al.} 2011, Science, 332,
  1396

\bibitem[{Beegle {et~al.}(1999)Beegle, Ajello, James, Dziczek, \&
  Alvarez}]{beegle1999high}
Beegle, L.~W., Ajello, J.~M., James, G.~K., Dziczek, D., \& Alvarez, M. 1999,
  Astronomy and Astrophysics, 347, 375

\bibitem[{Bertaux {et~al.}(1999)Bertaux, Costa, Mäkinen, Quémerais,
  Lallement, Kyrölä, \& Schmidt}]{bertaux_lyman-alpha_1999}
Bertaux, J.~L., Costa, J., Mäkinen, T., {et~al.} 1999, Planetary and Space
  Science, 47, 725, \dodoi{10.1016/S0032-0633(98)00130-5}

\bibitem[{{Bhatia} \& {Kastner}(1995)}]{Bhatia1995}
{Bhatia}, A.~K., \& {Kastner}, S.~O. 1995, \apjs, 96, 325,
  \dodoi{10.1086/192121}

\bibitem[{Biver \& Bockelée-Morvan(2019)}]{biver_complex_2019}
Biver, N., \& Bockelée-Morvan, D. 2019, ACS Earth and Space Chemistry,
  \dodoi{10.1021/acsearthspacechem.9b00130}

\bibitem[{Bockelée-Morvan {et~al.}(2016)Bockelée-Morvan, Crovisier, Erard,
  Capaccioni, Leyrat, Filacchione, Drossart, Encrenaz, Biver, de~Sanctis,
  Schmitt, Kührt, Capria, Combes, Combi, Fougere, Arnold, Fink, Ip,
  Migliorini, Piccioni, \& Tozzi}]{bockelee-morvan_evolution_2016}
Bockelée-Morvan, D., Crovisier, J., Erard, S., {et~al.} 2016, Monthly Notices
  of the Royal Astronomical Society, 462, S170, \dodoi{10.1093/mnras/stw2428}

\bibitem[{Bodewits {et~al.}(2016)Bodewits, Lara, A\textsc{\char13}Hearn,
  La~Forgia, Gicquel, Kovacs, Knollenberg, Lazzarin, Lin, Shi,
  {et~al.}}]{bodewits2016changes}
Bodewits, D., Lara, L.~M., A\textsc{\char13}Hearn, M.~F., {et~al.} 2016, \aj,
  152, 130

\bibitem[{Bodewits {et~al.}(2020)Bodewits, Noonan, Feldman, Bannister,
  Farnocchia, Harris, Li, Mandt, Parker, \& Xing}]{bodewits_carbon_2020}
Bodewits, D., Noonan, J.~W., Feldman, P.~D., {et~al.} 2020, Nature Astronomy,
  1, \dodoi{10.1038/s41550-020-1095-2}

\bibitem[{Calmonte {et~al.}(2016)Calmonte, Altwegg, Balsiger, Berthelier,
  Bieler, Cessateur, Dhooghe, Van~Dishoeck, Fiethe, Fuselier,
  {et~al.}}]{calmonte2016sulphur}
Calmonte, U., Altwegg, K., Balsiger, H., {et~al.} 2016, Monthly Notices of the
  Astronomical Society, 462, S253

\bibitem[{{Carlsson} \& {Judge}(1993)}]{Carlsson1993}
{Carlsson}, M., \& {Judge}, P.~G. 1993, \apj, 402, 344, \dodoi{10.1086/172138}

\bibitem[{Chaufray {et~al.}(2017)Chaufray, Bockelée-Morvan, Bertaux, Erard,
  Feldman, Capaccioni, Schindhelm, Leyrat, Parker, \&
  Filacchione}]{chaufray_rosetta_2017}
Chaufray, J.-Y., Bockelée-Morvan, D., Bertaux, J.-L., {et~al.} 2017, Monthly
  Notices of the Royal Astronomical Society, 469, S416

\bibitem[{Combi {et~al.}(2019)Combi, M{\"a}kinen, Bertaux, Qu{\'e}merais, \&
  Ferron}]{combi2019survey}
Combi, M., M{\"a}kinen, T.~T., Bertaux, J.-L., Qu{\'e}merais, E., \& Ferron, S.
  2019, Icarus, 317, 610

\bibitem[{Combi {et~al.}(1998)Combi, Brown, Feldman, Keller, Meier, \&
  Smyth}]{combi_hubble_1998}
Combi, M.~R., Brown, M.~E., Feldman, P.~D., {et~al.} 1998, The Astrophysical
  Journal, 494, 816, \dodoi{10.1086/305228}

\bibitem[{Combi \& Feldman(1992)}]{combi_iue_1992}
Combi, M.~R., \& Feldman, P.~D. 1992, Icarus, 97, 260,
  \dodoi{10.1016/0019-1035(92)90132-Q}

\bibitem[{Combi {et~al.}(2004)Combi, Harris, \& Smyth}]{combi2004gas}
Combi, M.~R., Harris, W.~M., \& Smyth, W.~H. 2004, Comets II, 1, 523

\bibitem[{Combi {et~al.}(2020)Combi, Mäkinen, Bertaux, Quémerais, Ferron, \&
  Coronel}]{combi_comet_2020}
Combi, M.~R., Mäkinen, T., Bertaux, J.-L., {et~al.} 2020, arXiv:2007.05138
  [astro-ph].
\newblock \url{http://arxiv.org/abs/2007.05138}

\bibitem[{Cunningham {et~al.}(2015)Cunningham, Spencer, Feldman, Strobel,
  France, \& Osterman}]{cunningham_detection_2015}
Cunningham, N.~J., Spencer, J.~R., Feldman, P.~D., {et~al.} 2015, Icarus, 254,
  178, \dodoi{10.1016/j.icarus.2015.03.021}

\bibitem[{Curdt {et~al.}(2001)Curdt, Brekke, Feldman, Wilhelm, Dwivedi,
  Sch{\"u}hle, \& Lemaire}]{curdt2001sumer}
Curdt, W., Brekke, P., Feldman, U., {et~al.} 2001, Astronomy \& Astrophysics,
  375, 591

\bibitem[{{Farnham} \& {Schleicher}(1998)}]{1998A&A...335L..50F}
{Farnham}, T.~L., \& {Schleicher}, D.~G. 1998, \aap, 335, L50

\bibitem[{{Feaga} {et~al.}(2015){Feaga}, {Feldman}, {A'Hearn}, {Bertaux},
  {Keeney}, {Knight}, {Noonan}, {Parker}, {Schindhelm}, {Steffl}, {Stern},
  {Vervack}, \& {Weaver}}]{2015DPS....4750306F}
{Feaga}, L.~M., {Feldman}, P.~D., {A'Hearn}, M.~F., {et~al.} 2015, in
  AAS/Division for Planetary Sciences Meeting Abstracts \#47, AAS/Division for
  Planetary Sciences Meeting Abstracts, 503.06

\bibitem[{{Feldman} {et~al.}(2004){Feldman}, {Cochran}, \&
  {Combi}}]{2004come.book..425F}
{Feldman}, P.~D., {Cochran}, A.~L., \& {Combi}, M.~R. 2004, {Spectroscopic
  investigations of fragment species in the coma}, ed. M.~C. {Festou}, H.~U.
  {Keller}, \& H.~A. {Weaver}, 425

\bibitem[{Feldman {et~al.}(2018)Feldman, Weaver, A’Hearn, Combi, \&
  Russo}]{feldman2018far}
Feldman, P.~D., Weaver, H.~A., A’Hearn, M.~F., Combi, M.~R., \& Russo, N.~D.
  2018, The Astronomical Journal, 155, 193

\bibitem[{Feldman {et~al.}(2015)Feldman, A'Hearn, Bertaux, Feaga, Parker,
  Schindhelm, Steffl, Stern, Weaver, Sierks, \&
  Vincent}]{feldman_measurements_2015}
Feldman, P.~D., A'Hearn, M.~F., Bertaux, J.-L., {et~al.} 2015, Astronomy \&
  Astrophysics, 583, A8, \dodoi{10.1051/0004-6361/201525925}

\bibitem[{Feldman {et~al.}(2016)Feldman, A'Hearn, Feaga, Bertaux, Noonan,
  Parker, Schindhelm, Steffl, Stern, \& Weaver}]{feldman_nature_2016}
Feldman, P.~D., A'Hearn, M.~F., Feaga, L.~M., {et~al.} 2016, The Astrophysical
  Journal, 825, L8, \dodoi{10.3847/2041-8205/825/1/L8}

\bibitem[{Feldman {et~al.}(2017)Feldman, A'Hearn, Bertaux, Feaga, Keeney,
  Knight, Noonan, Parker, Schindhelm, Steffl, Stern, Vervack, \&
  Weaver}]{feldman_fuv_2017}
Feldman, P.~D., A'Hearn, M.~F., Bertaux, J.-L., {et~al.} 2017, The Astronomical
  Journal, 155, 9, \dodoi{10.3847/1538-3881/aa9bf2}

\bibitem[{{Festou}(1981)}]{1981A&A....95...69F}
{Festou}, M.~C. 1981, \aap, 95, 69

\bibitem[{Fink \& Combi(2004)}]{fink2004effect}
Fink, U., \& Combi, M. 2004, Planetary and Space Science, 52, 573

\bibitem[{{Fink} {et~al.}(1998){Fink}, {Hicks}, {Fevig}, \&
  {Collins}}]{1998A&A...335L..37F}
{Fink}, U., {Hicks}, M.~D., {Fevig}, R.~A., \& {Collins}, J. 1998, \aap, 335,
  L37

\bibitem[{Green {et~al.}(2012)Green, Froning, Osterman, Ebbets, Heap,
  Leitherer, Linsky, Savage, Sembach, Michael~Shull, Siegmund, Snow, Spencer,
  Alan~Stern, Stocke, Welsh, Béland, Burgh, Danforth, France, Keeney, McPhate,
  Penton, Andrews, Brownsberger, Morse, \& Wilkinson}]{green_cosmic_2012}
Green, J.~C., Froning, C.~S., Osterman, S., {et~al.} 2012, The Astrophysical
  Journal, 744, 60, \dodoi{10.1088/0004-637X/744/1/60}

\bibitem[{{Groussin} {et~al.}(2007){Groussin}, {Hahn}, {Lamy}, {Gonczi}, \&
  {Valsecchi}}]{2007MNRAS.376.1399G}
{Groussin}, O., {Hahn}, G., {Lamy}, P.~L., {Gonczi}, R., \& {Valsecchi}, G.~B.
  2007, \mnras, 376, 1399, \dodoi{10.1111/j.1365-2966.2007.11553.x}

\bibitem[{{Groussin} \& {Lamy}(2003)}]{2003A&A...412..879G}
{Groussin}, O., \& {Lamy}, P. 2003, \aap, 412, 879,
  \dodoi{10.1051/0004-6361:20031496}

\bibitem[{Hall {et~al.}(1998)Hall, Feldman, McGrath, \& Strobel}]{hall1998far}
Hall, D., Feldman, P., McGrath, M.~A., \& Strobel, D. 1998, \apj, 499, 475

\bibitem[{Hansen {et~al.}(2016)Hansen, Altwegg, Berthelier, Bieler, Biver,
  Bockelée-Morvan, Calmonte, Capaccioni, Combi, De~Keyser, Fiethe, Fougere,
  Fuselier, Gasc, Gombosi, Huang, Le~Roy, Lee, Nilsson, Rubin, Shou, Snodgrass,
  Tenishev, Toth, Tzou, \& Simon~Wedlund}]{hansen_evolution_2016}
Hansen, K.~C., Altwegg, K., Berthelier, J.-J., {et~al.} 2016, Monthly Notices
  of the Royal Astronomical Society, 462, S491, \dodoi{10.1093/mnras/stw2413}

\bibitem[{{Haser}(1957)}]{Haser1957}
{Haser}, L. 1957, Bulletin de la Societe Royale des Sciences de Liege, 43, 740

\bibitem[{Huebner \& Mukherjee(2015)}]{huebner_photoionization_2015}
Huebner, W.~F., \& Mukherjee, J. 2015, Planetary and Space Science, 106, 11,
  \dodoi{10.1016/j.pss.2014.11.022}

\bibitem[{Jackson {et~al.}(1986)Jackson, Butterworth, \&
  Ballard}]{jackson_origin_1986}
Jackson, W.~M., Butterworth, P.~S., \& Ballard, D. 1986, The Astrophysical
  Journal, 304, 515, \dodoi{10.1086/164185}

\bibitem[{Kanik {et~al.}(2003)Kanik, Noren, Makarov, Vattipalle, Ajello, \&
  Shemansky}]{kanik_electron_2003}
Kanik, I., Noren, C., Makarov, O.~P., {et~al.} 2003, Journal of Geophysical
  Research: Planets, 108, 5126, \dodoi{10.1029/2000JE001423}

\bibitem[{Kelley {et~al.}(2013)Kelley, Lindler, Bodewits, A’Hearn, Lisse,
  Kolokolova, Kissel, \& Hermalyn}]{kelley2013distribution}
Kelley, M.~S., Lindler, D.~J., Bodewits, D., {et~al.} 2013, Icarus, 222, 634

\bibitem[{{Kobayashi} \& {Kawakita}(2010)}]{2010PASJ...62.1025K}
{Kobayashi}, H., \& {Kawakita}, H. 2010, \pasj, 62, 1025,
  \dodoi{10.1093/pasj/62.4.1025}

\bibitem[{Kramida {et~al.}(2019)Kramida, {Yu.~Ralchenko}, Reader, \& {and NIST
  ASD Team}}]{NIST_ASD}
Kramida, A., {Yu.~Ralchenko}, Reader, J., \& {and NIST ASD Team}. 2019, {NIST
  Atomic Spectra Database (ver. 5.7.1), [Online]. Available:
  {\tt{https://physics.nist.gov/asd}} [2017, April 9]. National Institute of
  Standards and Technology, Gaithersburg, MD.}

\bibitem[{{Lamy} {et~al.}(1998){Lamy}, {Toth}, {Jorda}, {Weaver}, \&
  {A'Hearn}}]{1998A&A...335L..25L}
{Lamy}, P.~L., {Toth}, I., {Jorda}, L., {Weaver}, H.~A., \& {A'Hearn}, M. 1998,
  \aap, 335, L25

\bibitem[{Lis {et~al.}(2019)Lis, Bockelée-Morvan, Güsten, Biver, Stutzki,
  Delorme, Durán, Wiesemeyer, \& Okada}]{lis_terrestrial_2019}
Lis, D.~C., Bockelée-Morvan, D., Güsten, R., {et~al.} 2019, Astronomy \&
  Astrophysics, 625, L5, \dodoi{10.1051/0004-6361/201935554}

\bibitem[{Lupu {et~al.}(2007)Lupu, Feldman, Weaver, \&
  Tozzi}]{lupu_fourth_2007}
Lupu, R.~E., Feldman, P.~D., Weaver, H.~A., \& Tozzi, G.-P. 2007, The
  Astrophysical Journal, 670, 1473, \dodoi{10.1086/522328}

\bibitem[{Makarov {et~al.}(2004)Makarov, Ajello, Vattipalle, Kanik, Festou, \&
  Bhardwaj}]{makarov_kinetic_2004}
Makarov, O.~P., Ajello, J.~M., Vattipalle, P., {et~al.} 2004, Journal of
  Geophysical Research: Space Physics, 109, A09303,
  \dodoi{10.1029/2002JA009353}

\bibitem[{Mandt {et~al.}(2016)Mandt, Eriksson, Edberg, Koenders, Broiles,
  Fuselier, Henri, Nemeth, Alho, Biver, {et~al.}}]{mandt2016rpc}
Mandt, K., Eriksson, A., Edberg, N., {et~al.} 2016, \mnras, 462, S9

\bibitem[{Mayyasi {et~al.}(2020)Mayyasi, Clarke, Combi, Fougere, Quemerais,
  Katushkina, Bhattacharyya, Crismani, Deighan, Jain,
  {et~al.}}]{mayyasi2020lyalpha}
Mayyasi, M., Clarke, J., Combi, M., {et~al.} 2020, The Astronomical Journal,
  160, 10

\bibitem[{McCoy {et~al.}(1992)McCoy, Meier, Keller, Opal, \&
  Carruthers}]{mccoy1992hydrogen}
McCoy, R., Meier, R., Keller, H., Opal, C., \& Carruthers, G. 1992, Astronomy
  and Astrophysics, 258, 555

\bibitem[{Meier \& A'Hearn(1997)}]{meier_atomic_1997}
Meier, R., \& A'Hearn, M.~F. 1997, Icarus, 125, 164,
  \dodoi{10.1006/icar.1996.5600}

\bibitem[{Miller-Ricci \& Uitenbroek(2002)}]{MillerRicci2002}
Miller-Ricci, E., \& Uitenbroek, H. 2002, The Astrophysical Journal, 566, 500,
  \dodoi{10.1086/337954}

\bibitem[{Morton \& Noreau(1994)}]{morton1994compilation}
Morton, D.~C., \& Noreau, L. 1994, The astrophysical journal supplement series,
  95, 301

\bibitem[{Protopapa {et~al.}(2018)Protopapa, Kelley, Yang, Bauer, Kolokolova,
  Woodward, Keane, \& Sunshine}]{protopapa_icy_2018}
Protopapa, S., Kelley, M. S.~P., Yang, B., {et~al.} 2018, The Astrophysical
  Journal, 862, L16, \dodoi{10.3847/2041-8213/aad33b}

\bibitem[{Protopapa {et~al.}(2014)Protopapa, Sunshine, Feaga, Kelley,
  A’Hearn, Farnham, Groussin, Besse, Merlin, \& Li}]{protopapa2014water}
Protopapa, S., Sunshine, J.~M., Feaga, L.~M., {et~al.} 2014, Icarus, 238, 191

\bibitem[{Pryor {et~al.}(2013)Pryor, Holsclaw, McClintock, Snow, Vervack,
  Gladstone, Stern, Retherford, \& Miles}]{pryor2013lyman}
Pryor, W.~R., Holsclaw, G.~M., McClintock, W.~E., {et~al.} 2013, in
  Cross-Calibration of Far UV Spectra of Solar System Objects and the
  Heliosphere (Springer), 163--175

\bibitem[{Rafelski(2018)}]{Rafelski2018}
Rafelski, M., e.~a. 2018, COS Data Handbook, Version 4.0, (Baltimore: STScI)

\bibitem[{Roettger {et~al.}(1989)Roettger, Feldman, A'Hearn, Festou, McFadden,
  \& Gilmozzi}]{roettger_iue_1989}
Roettger, E.~E., Feldman, P.~D., A'Hearn, M.~F., {et~al.} 1989, Icarus, 80,
  303, \dodoi{10.1016/0019-1035(89)90141-3}

\bibitem[{Roth {et~al.}(2017)Roth, Retherford, Ivchenko, Schlatter, Strobel,
  Becker, \& Grava}]{roth2017detection}
Roth, L., Retherford, K.~D., Ivchenko, N., {et~al.} 2017, The Astronomical
  Journal, 153, 67

\bibitem[{{Schulz} {et~al.}(1998){Schulz}, {Arpigny}, {Manfroid}, {Stuewe},
  {Tozzi}, {Cremonese}, {Rembor}, \& {Peschke}}]{1998A&A...335L..46S}
{Schulz}, R., {Arpigny}, C., {Manfroid}, J., {et~al.} 1998, \aap, 335, L46

\bibitem[{Sierks {et~al.}(2015)Sierks, Barbieri, Lamy, Rodrigo, Koschny,
  Rickman, Keller, Agarwal, A\textsc{\char13}Hearn, Angrilli,
  {et~al.}}]{sierks2015nucleus}
Sierks, H., Barbieri, C., Lamy, P.~L., {et~al.} 2015, Science, 347, aaa1044

\bibitem[{{Stern} {et~al.}(1998){Stern}, {Parker}, {Festou}, {A'Hearn},
  {Feldman}, {Schwehm}, {Schulz}, {Bertaux}, \& {Slater}}]{Stern1998}
{Stern}, S.~A., {Parker}, J.~W., {Festou}, M.~C., {et~al.} 1998, \aap, 335, L30

\bibitem[{Stern {et~al.}(2007)Stern, Slater, Scherrer, Stone, Versteeg,
  A\textsc{\char13}Hearn, Bertaux, Feldman, Festou, Parker,
  {et~al.}}]{stern2007alice}
Stern, S.~A., Slater, D., Scherrer, J., {et~al.} 2007, \ssr, 128, 507

\bibitem[{{Swings}(1941)}]{1941LicOB..19..131S}
{Swings}, P. 1941, Lick Observatory Bulletin, 508, 131,
  \dodoi{10.5479/ADS/bib/1941LicOB.19.131S}

\bibitem[{Weaver {et~al.}(2011)Weaver, Feldman, A'Hearn, Russo, \&
  Stern}]{weaver_carbon_2011}
Weaver, H.~A., Feldman, P.~D., A'Hearn, M.~F., Russo, N.~D., \& Stern, S.~A.
  2011, The Astrophysical Journal Letters, 734, L5,
  \dodoi{10.1088/2041-8205/734/1/L5}

\bibitem[{Weaver {et~al.}(1981)Weaver, Feldman, Festou, A'Hearn, \&
  Keller}]{weaver_iue_1981}
Weaver, H.~A., Feldman, P.~D., Festou, M.~C., A'Hearn, M.~F., \& Keller, H.~U.
  1981, Icarus, 47, 449, \dodoi{10.1016/0019-1035(81)90193-7}

\bibitem[{Woods {et~al.}(1998)Woods, Bailey, Eparvier, Lawrence, Lean,
  McClintock, Roble, Rottman, Solomon, Tobiska, {et~al.}}]{woods1998timed}
Woods, T.~N., Bailey, S.~M., Eparvier, F.~G., {et~al.} 1998, in Missions to the
  Sun II, Vol. 3442, International Society for Optics and Photonics, 180--191

\bibitem[{Woods {et~al.}(2000)Woods, Bailey, Eparvier, Lawrence, Lean,
  McClintock, Roble, Rottman, Solomon, Tobiska, {et~al.}}]{woods2000timed}
Woods, T.~N., Bailey, S., Eparvier, F., {et~al.} 2000, PHYS. CHEM. EARTH PART C
  SOL. TERR. PLANET. SCI., 25, 393

\bibitem[{Yang {et~al.}(2009)Yang, Jewitt, \& Bus}]{yang_comet_2009}
Yang, B., Jewitt, D., \& Bus, S.~J. 2009, The Astronomical Journal, 137, 4538,
  \dodoi{10.1088/0004-6256/137/5/4538}

\bibitem[{Yang {et~al.}(2014)Yang, Keane, Meech, Owen, \&
  Wainscoat}]{yang_multi-wavelength_2014}
Yang, B., Keane, J., Meech, K., Owen, T., \& Wainscoat, R. 2014, The
  Astrophysical Journal, 784, L23, \dodoi{10.1088/2041-8205/784/2/L23}

\end{thebibliography}

\end{CJK*}
\end{document}